\input amstex
\input amsppt.sty            
\magnification=1100		
\pagewidth{6.  truein}
\pageheight{9. truein}
\hoffset = .15in
\voffset = 0in
\parskip=.2in
\NoBlackBoxes
\TagsOnRight
\def\ol{\overline}
\def\q{\quad}

\def\hsk{\hskip-.25in}
\def\noin{\noindent}
\headline={\hfill \the \pageno}


\topmatter
\headline={\hfil \the \pageno}
\title   Self-Similarity in General Relativity \endtitle
\endtopmatter
\centerline{B. J. Carr}
\centerline{\it Astronomy Unit}
\centerline{\it Queen Mary and Westfield College}
\centerline{\it University of London}
\centerline{\it  London, England}
\smallbreak
\smallbreak
\centerline{and}
\smallbreak
\smallbreak
\centerline{A. A. Coley}
\centerline{\it Department of Mathematics}
\centerline{\it Statistics and Computing Science}
\centerline{\it Dalhousie University}
\centerline{\it Halifax, Nova Scotia}
\centerline{\it Canada B3H 3J5}
\bigbreak
\bigbreak

\abstract 
 The different kinds of self-similarity in general relativity are 
discussed, with special emphasis on similarity of the ``first'' kind,
corresponding to spacetimes admitting a homothetic vector. We then survey 
the various classes of self-similar solutions to Einstein's field equations 
and the different mathematical approaches used in studying them.
We focus mainly on spatially homogenous and spherically symmetric self-similar solutions, emphasizing their possible roles as asymptotic states for 
more general models. Perfect fluid spherically symmetric
similarity solutions have recently been completely classified,  and we discuss various astrophysical and cosmological applications
of such solutions. Finally we consider more general types of self-similar models.
\endabstract
\endtopmatter

\document
\baselineskip18pt

\heading{1. Introduction} \endheading

The purpose of this review is to summarize recent developments in the study of
self-similar solutions in general relativity and to discuss recent
applications of these solutions. It thus combines a mathematical and physical
approach to the subject. In these introductory remarks, we will first
discuss the issues involved in rather broad terms, this serving to
delineate the general scope of the review. We will then present some technical mathematical background to elucidate the nature of self-similarity. Finally, as preparation for the more detailed discussion in later sections, we will present an overview of t
he different types of solutions.

\subhead{A. Overview}\endsubhead 

\noin
{\it Forms of self-similarity} 

In Newtonian hydrodynamics self-similar solutions occur when the
physical quantities depend on functions of $x / l (t)$, where $x$ and
$t$ are independent space and time variables and $l$ is a
time-dependent scale.  This means that the spatial distribution of the
characteristics of motion remains similar to itself at all times during
the motion and that all dimensional constants entering the
initial and boundary conditions vanish or become infinite (Barenblatt
and Zeldovich 1972).  
When the form of the self-similar asymptotics can be
obtained from dimensional considerations, the solutions are referred to
as self-similar solutions of the first kind (Barenblatt and Zeldovich
1972). Examples of these appear in the study of strong  
explosions (Sedov 1946 and 1967, Taylor 1950) and  thermal waves
(Zeldovich and Kompaneets 1950, Barenblatt 1952, Zeldovich and Raizer
1963).  Otherwise, the solutions are referred to as self-similar solutions of the more general second kind.    
Self-similar solutions also describe the
``intermediate-asymptotic'' behaviour of solutions in the region in
which they no longer depend on the details of the initial and/or
boundary conditions but in which the system may still be far from
equilibrium.

In general relativity the concept of self-similarity is perhaps less straightforward, since in principle there are
various ways of generalizing the Newtonian concept and also a covariant characterization is required. 
First, it is important to distinguish between different types of
self-similarity. The existence of self-similar solutions of the {\it first} kind
is related to conservation laws and to the invariance of 
the problem with respect to the group of similarity transformations
of quantities with independent dimensions.  This can be characterized within 
general relativity by
the existence of a {\it homothetic} vector. In this case, one 
assumes a certain regularity of the limiting
process in passing from the original non-self-similar regime to the
self-similar regime.  However, in general
such a passage need not be regular, so the
expressions for the self-similar variables are not determined 
from dimensional analysis alone.  Solutions
are then called self-similar solutions of the {\it second} kind.
As in the Newtonian context, a characteristic of these solutions is that they   
contain {\it dimensional constants} which are not determined from the conservation 
laws but can be found by matching the self-similar
solutions with the non-self-similar solutions whose
asymptotes they represent (Barenblatt and Zeldovich 1972).  Most of this
review will be concerned with self-similarity of the first kind but we shall consider more general kinds of self-similarity in Section 5. In particular, the 
important example of kinematic self-similarity (Carter and Henriksen 1989,  Coley 1997a) will be reviewed.

Second, in general
relativity one must distinguish between
{\it geometrical} and {\it physical} self-similarity. Geometrical similarity is a property of the spacetime metric, whereas physical similarity is a property of the matter fields. As discussed in 
Section 1B, these need not be equivalent and the relationship between them also
depends on the nature of the matter. In much of this review we will be
focussing on perfect fluid solutions admitting a homothetic vector and in this
case geometrical self-similarity implies physical self-similarity. However, some of the discussion will pertain to more general fluids.

Third, it is important to distinguish between {\it continuous} and {\it discrete} self-similarity.  For example, in the spherically symmetric case, the continuous kind involves a similarity variable,
$\zeta$, with all dimensionless quantities $\Psi(\zeta)$ being invariant 
under any coordinate transformation for which $\zeta$ is constant. A discrete self-similarity is then one in which all dimensionless variables $\Psi$ repeat themselves on some spacetime scale: this condition can usually be written as $\Psi(\tau, \zeta) = 
\Psi(\tau -n \Delta, \zeta)$ for some constant $\Delta$, where n is an integer and $\tau$ is another variable. Since one recovers continuous self-similarity in the limit $\Delta\rightarrow 0$, continuous self-similarity can
be regarded as a special case of 
discrete self-similarity and is much easier to deal with mathematically. 
Although we are mainly 
concerned with continuous self-similarity in this review, discrete self-similarity is of great interest in its own right and, as we will see in
Section 4C, a focus of considerable interest in the context of critical phenomena.

\noin
{\it Relevance of self-similarity} 

There are two important reasons for studying self-similar solutions
of the Einstein field equations (EFEs). First, the assumption of self-similarity reduces the
mathematical complexity of the governing differential equations, often leading in problems of physical interest to the reduction of partial differential equations 
(PDEs) to ordinary differential equations (ODEs). This makes such solutions
easier to study mathematically. Indeed self-similarity in the broadest (Lie) sense refers to an invariance which allows such a reduction.

Second, self-similar solutions play an important role in describing the asymptotic properties of more general models. This is discussed in detail in Section 2B for spatially homogeneous models and 
the same idea may apply in some spherically symmetric contexts. For example, the expansion of the Universe from the big
bang and the collapse of a star to a singularity might both
exhibit self-similarity in some form since it might be expected
that the initial conditions would be ``forgotten'' in some sense. In the cosmological context, the suggestion that fluctuations might naturally evolve from complex initial conditions via the Einstein equations to self-similar form  has been termed the ``s
imilarity hypothesis'' (Carr 1993). This certainly does not apply in all circumstances but it may do so 
whenever one has non-linear perturbations and non-zero pressure. One of our aims here is to discuss under what circumstances the 
similarity hypothesis might hold. As a first step in this direction, 
we discuss the stability of self-similar solutions to non-self-similar
perturbations in Section 4D.

\noin
{\it Classes of self-similar solutions}

As discussed in Section 1C, there are many different classes of
self-similar solutions. In particular, there are self-similar spatially homogeneous models, which will be reviewed in Section 2, and self-similar spherically symmetric models, which have now been classified completely 
(Carr and Coley 1998a) and will be reviewed in Section 3. There are other exact 
homothetic models, including for example self-similar $G_2$ models and plane-symmetric models, although
these are discussed in less detail.

These different types of solutions tend to attract different types of mathematical analysis. Recent studies of the spatially homogenous models often use a dynamical systems approach. This is because the governing ODEs reduce to an autonomous system and th
is approach facilitates the qualitative
analysis of the models. As discussed in Section 3A, some studies of the spherically symmetric models have also used a dynamical systems approach.
However, because of the mathematical simplicity involved in this case, one can often write the solutions explicitly, as emphasized in Section 3C, and this may offer a more physically intuitive approach. 

Self-similar models can also be analyzed using either a tetrad or coordinate approach, a variety of preferred gauges (e.g., coordinate systems), and a number of natural variables.
In the spherically symmetric case, in particular, one can use three possible  coordinate approaches. The first one uses ``comoving'' coordinates and was the one pioneered by Cahill and Taub (1973) and then followed by Carr and Henriksen and coworkers. The
 second approach, followed by  
Bogoyavlenski and coworkers, uses ``homothetic'' coordinates, in which the homothetic vector is along either the time or space axis. In this case, the equations can be reduced to that of a dynamical system and one can exploit
results derived from the study of hypersurface homogeneous models. A third approach uses ``Schwarzschild'' coordinates and was included in the analysis of Ori and Piran (1990). In Section 3 we will mainly emphasize the first approach.

 \noin
{\it Applications of self-similarity} 

Besides their intrinsic mathematical interest, there are many applications
of similarity solutions in astrophysics and cosmology. 
The astrophysical applications include gravitational collapse and the occurrence of naked singularities (Section 4A). Indeed, most of the examples of naked
singularities in the literature involve self-similar solutions. The cosmological applications include features of gravitational clustering and cosmic voids (Section 4B). These features are particularly relevant because they allow the 
similarity hypothesis to be tested observationally (Section 4D).  A 
distinctively relativistic application includes the crucial role of self-similar solutions in critical phenomena (Section 4C), surely one of the most exciting 
developments in general relativity
in recent years.

We hope that the discussion in Section 4 will prove useful in drawing connections between different areas of research. We do not pretend that our selection of applications is complete - nearly all of them are drawn from
the spherically symmetric context - but it should
be broad enough to give a taste of the subject. Also we will be discussing topics which involve  different areas of expertise, so we will attempt to 
avoid too many technicalities.  Note that we shall restrict our attention to 
general relativity in this review, although there are many applications
of self-similar solutions in other theories of gravity.

 \subhead{B.  Mathematical Background}\endsubhead

In this review we shall be particularly concerned with the case in which the
source of the gravitational field  is a perfect fluid; i.e.,
the energy-momentum tensor is given by
$$T_{ab} = (\mu + p) u_a u_b + pg_{ab}, \tag1.1$$
where $u^a$ is the normalized fluid $4$-velocity, $\mu$ is the density and $p$ is the pressure. Unless  stated otherwise, a linear barotropic equation of 
state of the form
$$p = \alpha \mu \tag1.2 $$
will be assumed, where the constant $\alpha$ satisfies $0 \leq \alpha \leq 1$ for
ordinary
matter and $-1 \leq \alpha < -1/3$ for models
that undergo inflation.  Causality requires $-1 \leq \alpha \leq 1$.

Cahill and Taub (1971) were the first to study perfect fluid similarity solutions
in general relativity.  They did so in the cosmological context under the 
assumption of spherical symmetry. They assumed that  all dependent variables are   functions of a single {\it dimensionless} combination of  
$r$ and $t$  (i.e., the solution is invariant under the
transformation $\overline{t} = at$, $\overline{r} = ar$ for any constant $a$) and that
the model contains no
other dimensional constants.  This corresponds to the existence of 
a similarity  of the {\it first kind} and they showed that it can be 
invariantly 
formulated in terms of the existence of a {\it homothetic vector}.  

For a general spacetime a proper homothetic
vector (HV) is a vector field $\pmb{\xi}$ which satisfies  
$${\Cal L}_\xi g_{\mu v} = 2g_{\mu v}, \tag1.3$$
where $g_{\mu v}$ is the metric and ${\Cal L}$ denotes Lie differentiation along $\pmb{\xi}$.  An arbitrary constant on the right-hand-side of (1.3) has been rescaled to unity.  If this constant is
zero, i.e., ${\Cal L}_\xi g_{u v}=0$, then $\pmb{\xi}$ is a Killing vector (KV).  A homothetic motion 
or homothety captures the geometric notion of ``invariance under scale 
transformations''. From (1.3) it follows that
$${\Cal L}_\xi R^a \,_{  bcd} = 0, \tag1.4  $$
and hence 
$${\Cal L}_\xi R_{ab} = 0, \enskip {\Cal L}_\xi G_{ab} = 0. \tag 1.5a,b $$
A vector field $\pmb{\xi}$ that satisfies equation (1.4) is called a curvature 
collineation,
one that satisfies equation (1.5a) is called a Ricci collineation, and one
that satisfies equation (1.5b) is called a matter collineation.

For a perfect fluid, it follows from equation (1.3) and the EFEs that the physical quantities transform according
to
$${\Cal L}_\xi u^a = -u^a, \tag1.6 $$
and
$${\Cal L}_\xi \mu = - 2 \mu, \enskip {\Cal L}_\xi p = -2p, \tag1.7 $$
where
$${\Cal L}_\xi T_{ab} = 0 \tag1.8 $$ 
(Cahill and Taub 1971, Eardley 1974). Perfect fluid spacetimes admitting a HV within general relativity
have been comprehensively studied by Eardley (1974). In such 
spacetimes  equations (1.6) and (1.7) imply that all physical quantities transform according to their
respective dimensions, so   ``geometrical''  
and ``physical'' self-similarity coincide.  However, this need
not always be the case, and it is unfortunate that a rather misleading terminology has been introduced (Eardley 1974) which equates self-similarity
with the existence of a homothety and which refers to the homothetic group as the similarity group.

The question of whether the matter field exhibits the same symmetries
as the geometry within general relativity is called the symmetry
``inheritance'' problem.  If the source is not a perfect fluid, then the spacetime symmetries  
need not be inherited by the matter
(Coley and Tupper 1989), so a homothety is a purely geometric property of a spacetime rather than 
a self-similarity.  In this case, it is only through the EFEs
that properties of the matter like similarity can be inferred. On the other hand, if the matter fields exhibit self-similarity, then
the EFEs  place restrictions on the geometry.  For example,
if the self-similarity is of the first kind (i.e., resulting from dimensional considerations), then ${\Cal L}_\xi T_{ab} = 0$ implies
$${\Cal L}_\xi G_{ab}= 0, \tag1.9 $$
in which case $\pmb{\xi}$ is a matter collineation (Kramer et al. 1980, Carot et al. 1994).  Although a HV satisfies equation (1.9), a matter 
collineation
is not necessarily a HV.  Indeed, equation (1.9) need not
imply equation (1.5a), which in turn need not imply
equation (1.4); i.e., neither a curvature collineation nor even a Ricci collineation need be a HV.  
The general problem of determining the constraints 
on the form of the metric from an equation 
like (1.9) 
has been termed the ``inverse'' symmetry inheritance problem, and the
study of matter collineations in which ${\Cal L}_\xi C^a \,_{bcd} \neq 0$ 
(otherwise $\pmb{\xi}$ is necessarily a HV) was recently undertaken by 
Carot et al. (1994).

\subhead{C. Spacetimes Admitting a Homothetic Vector}\endsubhead

The differential geometric properties of HVs were studied by Yano (1955).  The 
totality of HVs on a spacetime form a Lie algebra $H_n$ of dimension $n$ 
which  (if $H_n$ is non-trivial) contains an $(n-1)$ dimensional 
subalgebra of KVs, 
$G_{n-1}$.  Except when the spacetime is conformal to a ``generalized 
plane-wave'' spacetime , it follows that if the orbits of $H_n$ are $r$-dimensional, then the orbits of $G_{n-1}$ are 
$(r-1)$-dimensional (Eardley 1974). If, in addition,  the  spacetime is   not conformally flat, then
it is conformally related to a spacetime for which the
Lie algebra $H_n$ is the Lie algebra of KVs (Defrise-Carter 1975).  
In the trivial case of a (locally) flat spacetime, the dimension of 
the homothetic algebra is eleven and that of its associated Killing
subalgebra is ten.  The orbits of the homothetic group and the
isometry group can coincide only if they are four-dimensional 
or  three-dimensional and null, the resulting spacetime is consequently 
either locally flat or is a special type of ``generalized plane-wave''
 spacetime (cf. Hall and Steele 1990).

Vacuum spacetimes admitting a HV were studied by McIntosh (1975), who showed that a non-flat vacuum spacetime can only admit
a non-trivial HV if that HV is neither null nor hypersurface-orthogonal.
He also showed that a perfect 
fluid spacetime cannot admit a non-trivial
HV which is orthogonal to the fluid $4$-velocity unless $p = \mu$.  If a 
spacetime 
admits a non-trivial HV and there is an equation of state of the form
$p = p(\mu)$, then necessarily $p = \alpha \mu$ (Cahill and Taub 1971,  Wainwright 1985); 
i.e., equation (1.2) 
results from equation (1.7).  We note that homothetic initial data is 
preserved by the EFEs 
(Eardley 1974).  In the case of radiation (with $p = \frac{1}{3} \mu$),   
$T \equiv T^{ab} g_{ab} = 0$ and
the existence of a HV implies the existence of a conserved current.
In general, if we define  $P^a = T^{ab} \xi_b$,   
energy-momentum 
conservation implies $P^a \,_{;a} = T$.  For radiation, 
$P^a = \frac{\mu}{3}( 4 u^a u_b \xi^b + \xi^a)$ and so $P^a \,_{;a} = 0$.

In addition to flat Minkowski spacetime, all FRW models admit a timelike HV
in the special case of  matter with $p = - \frac{1}{3} \mu$ (Eardley 1974).  
Otherwise, 
only the flat  model admits a HV, and this occurs for all $p = \alpha \mu$ 
models in which the scale function has power-law dependence
on time (Maartens and Maharaj 1986).

There are many exact self-similar spatially homogeneous and spherically symmetric
solutions, and these will be reviewed in Sections 2 and 3, respectively.  In 
addition, there are a number of exact homothetic $G_2$ 
models.  In these solutions there are two commuting spacelike KVs
acting orthogonally transitively, which together with the HVs form a
three-dimensional homothety group, $H_3$.  Exact solutions have been found
for the cases in which the orbits of $H_3$ are spacelike (Eardley 1974, Luminet 1978, Chao 1981, 
Carot and Sintes 1997), timelike (Hewitt and Wainwright 1990, Hewitt,
Wainwright and Goode, 1988, Hewitt, Wainwright and 
Glaum 1991, Uggla 1992, Carot and Sintes
1997), and  null (Carot and Sintes 1997), although the emphasis
in the timelike case has primarily been on the qualitative analysis of the models (cf. Wainwright and Ellis 1996).  Recently, Haager and Mars (1998) have 
analyzed algebraically general, non-diagonal $G_2$ self-similar tangent 
dust models (which are tangent in the sense that the fluid flow is tangent to the orbits of the $H_3$).
Homothetic cylindrically symmetric perfect fluid solutions also exist, 
but to our knowledge these only occur as special cases of known (more 
general) 
abelian $G_2$ solutions.  Abelian $G_2$ models in 
which one of the commuting KVs is timelike (i.e., the stationary
axisymmetric case) are also of astrophysical interest (Kramer et al. 1980).
Finally, exact self-similar solutions have been found for
plane symmetric spacetimes (Taub 1972, Shikin 1979, Chao 1981, Foglizzo and Henriksen 1993), 
hyperbolically symmetric spacetimes (Chao 1981),
Weyl spacetimes (Godfrey 1972),
and  diagonal hypersurface homogeneous spacetimes (Uggla et al. 1995).
Carot and Sintes (1997) have recently studied spacetimes 
admitting an $H_3$, in which the two (spacelike)
KVs are not necessarily orthogonally transitive nor commuting 
(and in which the perfect fluid does not
necessarily admit of a barotropic equation of state), or 
an $H_4$.  This extends and unifies previous work cited above.

\heading{2. Self-Similar models as asymptotic states of more general models   }\endheading

Self-similar models are often 
related to the asymptotic states of more general models (Hsu and Wainwright 1986). In particular, self-similar
models play an important role in the asymptotic
properties of spatially homogeneous models, spherically
symmetric models, $G_2$ models and silent universe
models (Bruni et al. 1995). In this section, we will focus on
spatially homogeneous models, which have been discussed in Ellis and 
MacCallum (1969) and Kramer et al. (1980), and $G_2$ models, which have been discussed in Kramer et al. (1980). The terminology used follows 
that of these references.  For the definitions of any technical terms in dynamical system theory used below, reference should be made to 
any modern textbook or  Wainwright and Ellis (1997;
WE).  We note that the self-similar 
Bianchi models discussed below are {\it transitively} self-similar, in the sense
that the
orbits of the $H_4$ are the whole spacetime, while the self-similar $G_2$ 
 and  spherically symmetric models are not transitively 
self-similar (unless they admit additional symmetry).
However, the three exact power-law self-similar spherically symmetric
solutions discussed in Section 3C
are  transitively self-similar.

\subhead{A. Spatially Homogeneous Models}\endsubhead

Many people have studied self-similar spatially homogeneous models, 
both as exact solutions and in the context of 
qualitative analyses (see WE and Coley 1997b and references therein).
Exact spatially homogeneous solutions were first displayed in 
early papers; however, 
it was not until after 1985 that many of them were recognized by Wainwright
(1985) 
and Rosquist and Jantzen (1985) as being self-similar
[although Eardley (1974) first pointed out  that some simple
Bianchi models are self-similar and appears to have been
the first to have introduced the notion of asymptotic self-similarity in 
cosmology].  The complete set of  
 self-similar orthogonal spatially
homogeneous perfect fluid and vacuum solutions were  given   
by Hsu and Wainwright (1986) and they have also been reviewed in WE.  
Kantowski-Sachs models were studied by Collins (1977).  Exact self-similar
solutions   
were given by Burd and Coley (1994)  and Coley and van den 
Hoogen (1994a, 1995) for imperfect fluid sources and by Feinstein and Iba\~nez (1993), Iba\~nez et al.
(1993) and Coley et al. (1997) for scalar field models with an 
exponential potential.

Spatially homogeneous models have attracted considerable
attention since the governing equations reduce to a relatively simple 
finite-dimensional
dynamical system, thereby enabling the models
to be studied by standard qualitative techniques.  Planar systems were initially analyzed by
Collins (1971, 1974) and a comprehensive study of general
Bianchi models was made by Bogoyavlenski and Novikov (1973) and
Bogoyavlenski (1985) and more recently (using automorphism variables and
Hamiltonian techniques) by Jantzen and Rosquist (Jantzen 1984, Rosquist 1984,
Jantzen and Rosquist 1986, Rosquist and Jantzen 1988, Rosquist et al. 1990).
Perhaps the most illuminating approach has been that of
Wainwright and collaborators (Hsu and Wainwright 1986,
Wainwright and Hsu 1989, Hewitt and Wainwright 1993), 
in which the more physically or geometrically natural 
expansion-normalized (dimensionless) configuration variables are 
used. In this case, the
physically admissible states typically lie within a bounded region,
the dynamical system remains analytic both in the physical
region and its boundaries, and the asymptotic
states typically lie on the boundary represented by exact physical
solutions rather than having singular behaviour.
We note that  the 
physically admissible states do not lie in a bounded region for Bianchi 
models of types VII$_0$, VIII and IX;
see WE for details.

Wainwright utilizes the orthonormal frame method
(Ellis and MacCallum 1969) and introduces expansion-normalized
(commutation function) variables and a new ``dimensionless'' time 
variable to study spatially
homogeneous perfect fluid models satisfying $p = \alpha \mu$.    The
equations governing the models form an $N$-dimensional
system of coupled autonomous ODEs.  When the ODEs
are written in expansion-normalized
variables, they admit a symmetry which allows the equation for the
time evolution of the expansion $\theta$  (the Raychaudhuri equation) to decouple.  The reduced $N-1$-dimensional dynamical system is then
studied.  At all of the singular points of the reduced
system,  $\dot{\theta}$ is proportional to $\theta^2$  and hence all such
points correspond to transitively self-similar cosmological models (Hsu
and Wainwright 1986).  This is why the   self-similar models play an important 
role in describing the asymptotic dynamics of the Bianchi models.

For orthogonal Bianchi models of class A, the resulting reduced 
state space is five-dimensional (Wainwright and Hsu 1989). Orthogonal Bianchi
cosmologies of class B were studied
by Hewitt and Wainwright (1993) and are governed by
a five-dimensional system of analytic ODEs with constraints.
In further work, imperfect fluid Bianchi models were
studied under the assumption that all physical quantities
satisfy ``dimensionless equations of state'', thereby 
ensuring that the singular points of the resulting reduced
dynamical system are represented by exact self-similar solutions
(Coley and van den Hoogen 1994a and b).  Models satisfying the linear
Eckart theory of irreversible thermodynamics were studied by Burd and 
Coley (1993) 
 and Coley and van den Hoogen (1994a), those satisfying the truncated 
causal theory 
of Israel-Stewart by Coley and van den Hoogen (1995),
and those satisfying the full (i.e., non-truncated) relativistic 
Israel-Stewart 
theory by Coley et al. (1996).   The singular points of 
the reduced dynamical system for scalar
field Bianchi cosmological models with an exponential potential again 
correspond to exact self-similar solutions; such models have been 
studied by Iba\~nez et al. (1995), van den Hoogen et al. (1997) 
and Coley et al. (1997).

It is interesting to ask under what circumstances the
singular points correspond to (and hence the
asymptotic properties of Bianchi models can be represented by)
self-similar models.  This depends critically on the equation
of state.  For perfect fluid models with $p/\mu$ 
asymptotically constant, it is plausible that self-similarity
of the asymptotic limits is preserved (Wainwright and Hsu 1989)
and this was indeed proved for a class of two-fluid models
(Coley and Wainwright 1992).  However, this property is not
robust, and self-similarity of the asymptotic limits is broken for 
perfect fluid models with
more complicated
equations of state or for imperfect fluid models that do not have ``dimensionless
equations of state'' (Coley and van den Hoogen 1994b).
It is also broken for sources consisting of a homogeneous  scalar field
with a non-exponential potential (cf. Ibanez et al. 1995) or if the strong energy condition is violated (e.g., if there
is a cosmological constant).

\subhead{B. Self-Similar Models as Asymptotic States of Bianchi Models}\endsubhead

We now discuss the primary role of exact self-similar models in describing
the asymptotic states of Bianchi models, again assuming $p = \alpha \mu$ with $0 \leq \alpha \leq 1$.  We will summarize the work of Wainwright and Hsu (1989)
and Hewitt and Wainwright (1993), who studied the asymptotic states of orthogonal spatially homogeneous models in terms of attractors of the 
associated dynamical system for class $A$ and class $B$ models, respectively.  Due to the existence of monotone functions, it is known
that there are no periodic or recurrent orbits in class A models.
Although ``typical'' results can be proved in a number 
of Bianchi type B cases, these are not ``generic'' due to the lack of knowledge of appropriate monotone functions.  In particular, there are no sources or sinks in the Bianchi invariant sets \newline $B^{\pm}_\alpha$ (VIII) or $B^\pm$(IX).

\item{$\bullet$} A large class of orthogonal spatially homogeneous models (including all class B models) are asymptotically self-similar
at the initial singularity and are approximated by exact perfect fluid or 
vacuum self-similar
power law models.  Examples include
self-similar Kasner vacuum models or self-similar locally rotationally
symmetric (class III) Bianchi type II perfect fluid models (Collins and Stewart 1971; see also Collins 1971 and Doreshkevich et al. 1973).

\noindent
However, this behaviour is not generic; general orthogonal models of Bianchi type IX and VIII 
have an oscillatory behaviour with chaotic-like characteristics, 
with the matter density becoming dynamically negligible as one follows the evolution into the past towards the initial singularity.  Ma and Wainwright (1994) 
show that the orbits of the associated cosmological 
dynamical system are negatively asymptotic to a lower two--dimensional 
attractor. 
This is the union of three ellipsoids in ${\Bbb R}^5$ consisting 
of the Kasner ring joined by
Taub separatrices; the orbits spend most of the time near the 
Kasner vacuum 
equilibrium points.  Clearly  the self-similar Kasner models play a 
primary role in the 
asymptotic behaviour of these models.

\item{$\bullet$}  Exact self-similar power law models can also 
approximate general Bianchi models at intermediate 
stages of their evolution  (e.g., radiation Bianchi VII$_h$ models; 
Doreshkevich et al., 1973).  Of special interest are those models which can be 
approximated by an isotropic solution at an intermediate stage of their evolution (e.g., those models whose orbits spend a 
period of time near to a flat Friedmann equilibrium point).

\noindent
This last point is of particular importance in relating Bianchi models to the 
real universe, and is discussed further in general terms in WE (see, especially, Chapter 15) and specifically for Bianchi VII$_h$ models in Wainwright et al. (1998).  In particular, the flat Friedmann equilibrium point is 
universal in that it is contained in the state space of each Bianchi type.  Isotropic intermediate behaviour has also been found in tilted Bianchi V models (Hewitt and Wainwright 1992), and it appears that many tilted models have isotropic intermediate be
haviour (see WE).

\item{$\bullet$}  Self-similar solutions can describe the behaviour 
of Bianchi models at late times (i.e., as $t \rightarrow \infty$). 
Examples include self-similar flat space and self-similar homogeneous 
vacuum plane waves (Collins 1971, Wainwright 1985).

\noindent
All models expand indefinitely except for the Bianchi IX models.  The 
question of which Bianchi models can isotropize was 
addressed in the famous paper by Collins and Hawking (1973), in which it was shown that, for physically reasonable matter, the set of homogeneous initial data that give rise to models that isotropize asymptotically to the future is of zero 
measure in the space of all homogeneous initial data (see also Barrow and Tipler 1986, and WE).

\noindent
All vacuum models of Bianchi (B) types IV, V, VI$_h$ and 
(especially) VII$_h$ are asymptotic to plane wave states to the future.
Type V models tend to the Milne form of flat spacetime (Hewitt and Wainwright 1993). Typically, and perhaps generically (Hewitt and
Wainwright 1993), non-vacuum models are asymptotic in the future
to either plane-wave vacuum solutions (Doroshkevich et al. 1973, 
Siklos 1981) or non-vacuum Collins type VI$_h$ solutions (Collins 1971).

\noindent
Bianchi (A) models of types VII$_o$ (non-vacuum) and VIII 
expand indefinitely but are found to have oscillatory (though 
non-chaotic) behaviour in the Weyl curvature (Wainwright, unpublished).
Bianchi type IX models obey the ``closed universe recollapse'' conjecture
(Lin and Wald 1989).  All orbits in the Bianchi invariant 
sets $B^\pm_\alpha(VII_0$) ($\Omega > 0), B^\pm_\alpha(VIII)$ and $B^\pm(IX)$ are 
positively departing; in order to analyse the
future asymptotic states of such models, it is necessary to compactify phase-space.  The description of these models in terms
of conventional expansion-normalized variables is only valid up to the point of 
maximum expansion (where $\theta = 0)$, although recently Wainwright has 
introduced  more appropriate variables which are valid for all values of $\theta$ (WE).  

In summary, due to the non-existence of periodic, recurrent and homoclinic orbits in the Bianchi
state space (deduced from the existence of monotone functions), 
the dynamical behaviour of Bianchi models is dominated by equilibrium points
and heteroclinic sequences (or heteroclinic cycles contained
in the Mixmaster attractor for class A models).
This is why self-similar models, which correspond
to equilibrium points, play a dominant role in the dynamics
of Bianchi cosmological models. These issues are further
discussed in WE.  In particular, one can
generalize  the above analysis to the exceptional
Bianchi VI$_{-1/9}$ models, to two-fluid models (Coley and
Wainwright 1992), and to inflationary models with
$-1 \leq \alpha < -1/3$ (cf. the cosmic no-hair theorems;  
Wald 1983).  Tilted Bianchi models and models with more
general sources than a perfect fluid (including, for example,
scalar fields, imperfect fluids and magnetic fields) are also
discussed in WE (see, especially, Chapter 8
by Hewitt, Uggla and Wainwright).  Self-similar spatially homogeneous
massless scalar
field models, which are formally equivalent to stiff $(\alpha =1)$ perfect
fluid models, have also been discussed by Coley and Wainwright (1998).

\subhead{C. $G_2$ Models}\endsubhead

Inhomogeneous perfect fluid $p = \alpha \mu$    
cosmological models admitting two commuting spacelike KVs acting orthogonally
transitively -- the so-called $G_2$ cosmologies --- have been studied
by Hewitt and Wainwright (1990) with a view to describing
their asymptotic behaviour near  the big bang and at late times.
In particular, they showed
that the EFEs can be written as an autonomous system of
first-order, quasi-linear (formally hyperbolic) PDEs (without
constraints) in terms of two independent dimensionless
variables, the state space being an infinite-dimensional
function space.  By defining the dynamical equilibrium
states in terms of an appropriately invariantly defined time
derivative of the state vector being zero, Hewitt and Wainwright (1990)
prove that these states correspond to cosmological models that are
self-similar (but not necessarily spatially homogeneous).  In this case, 
the EFEs reduce to a system of autonomous ODEs
(with spatial dependence) that can be studied qualitatively
by normal techniques; the spatially homogeneous subcases
have been studied previously (see above).

Hewitt and Wainwright (1990) conjecture that the dynamical equilibrium states
may describe the asymptotic or intermediate dynamical
behaviour of the orthogonally transitive $G_2$ models.  In particular,
they show that in the subclass of separable diagonal $G_2$ cosmologies
(in which the two KVs are hypersurface orthogonal), the models do
indeed asymptote towards the dynamical equilibrium states.  Thus the models in this special 
subclass are asymptotically self-similar.  In these models, the
orbits of the three-dimensional homothetic group are timelike,
the velocity vector being tangent to the group orbits, so the time evolution
is completely determined and the spatial structure is
governed by a two-dimensional plane autonomous system
of ODEs. The qualitative properties
of the diagonal self-similar $G_2$ cosmologies have been studied by Hewitt et al. (1988, 1991).

It remains to be determined to what extent a typical $G_2$ cosmology
which expands indefinitely from an initial singularity is asymptotically 
self-similar into the past
and the future.  Since the flat Friedmann model is a saddle point of the 
governing $G_2$ evolution equations, intermediate isotropization will 
occur for a subset of models, but the size of this subset 
of $G_2$ models is unclear.  These issues are discussed further in WE.

\heading{3. Spherically Symmetric Models}\endheading

The most extensive literature on self-similarity involves spherically
symmetric models since these obviously afford the greatest mathematical
simplification and have a number of important applications. The first work
focussed on dust ($\alpha =0$) solutions
since, in
this case, the solutions can often be expressed analytically and are just a
special subclass of the more general spherically symmetric
Tolman-Bondi solutions (Tolman 1934, Bondi 1947, Bonnor 1956); see, 
for example, Gurovich (1967), Dyer (1979), Chao (1981), Maharaj (1988) and, 
more recently, Joshi and
Darivedi (1993, Sintes (1996) and Carr and Coley (1998a). More extensive references can
be found in Kramer et al. (1980)
and Krasinski (1997). Self-similar dust solutions have played 
an important historical role in the subject but here we will mainly focus
on the more general case in which the fluid has an equation of state
$p=\alpha\mu$ with $0<\alpha<1$. It has recently been claimed that such
models can be classified completely, which makes our discussion
particularly timely. Much of the analysis will also be applicable in the
dust ($\alpha=0$) and stiff fluid ($\alpha =1$) limits but it should be
cautioned that not all features of the solutions will carry over in these
special cases. In some contexts, we will consider negative pressure models.
Here we require  $-1<\alpha<0$ but it should be noted that $\alpha =-1$
and $\alpha =-1/3$ are also special cases.

\subhead{A. Different Mathematical Approaches}\endsubhead

Due to the existence of several preferred geometric structures in self-similar
spherically symmetric perfect fluid models, a number of different 
approaches (i.e., coordinate systems) may be used in studying them
(Bogoyavlenski
1985). In particular, one can use ``comoving'' coordinates, ``homothetic''
coordinates or ``Schwarzschild'' coordinates. All of these approaches are 
complementary and which is most suitable depends on what type of problem is
being studied. The relationship between the various approaches, and
the precise coordinate transformations between them, can be found in a number of
sources (see Section IV.3 in Bogoyavlenski 1985, Ori and Piran 1990, Appendix
C of Carr and Coley 1998a, Appendices B of Goliath et al. 
1998a  and 1998b, and Carr et al. 1998).

In the comoving approach, pioneered by Cahill and Taub (1973) and employed
by Carr and Henriksen and coworkers and Ori and Piran (1990), the
coordinates
are adapted to the fluid four-velocity vector. We shall primarily adopt
this approach here since it affords
the best physical insights and is the most convenient one with which to
discuss the solutions explicitly. Even within
this approach, different authors use different notation, so it is sometimes
difficult to relate their results; in the discussion below we will
primarily use the notation of Cahill and Taub. Recently Carr and Coley
(1998a) presented  
a comprehensive and unified analysis of spherically symmetric 
self-similar perfect fluid models using the comoving approach, relating
many of the results obtained earlier by Ori and Piran (1990)
and Foglizzo and Henriksen (1993).

In the homothetic approach, used by  
Bogoyavlenski and coworkers and adopted more recently by Brady (1994) and
Goliath et al. (1998a and 1998b), the coordinates are adapted to the homothetic vector
and a ``conformally static'' metric is employed. In this case, the governing
equations reduce to an autonomous system of ODEs and hence dynamical systems
theory can be exploited to study them. The results of the dynamical systems
analysis complement and, in some cases, provide more rigorous
demonstrations of the results obtained in the comoving approach. However,
in the homothetic
approach spacetime must be covered by several coordinate patches, one in which
the HV is spacelike and one in which it is timelike. These
regions must then be joined by a surface in which the HV is
null and this surface is associated with important physics. 
Bogoyavlenski (1985) studied the spacelike and timelike cases 
simultaneously 
and continuously matched the two regions to obtain the behaviour of solutions
crossing the null surface; however, it should be noted that Bogoyavlenski 
changed to comoving coordinates explicitly in order to describe the physics
of the associated solutions. 

Recently Goliath et al. (1998a and 1998b) have reinvestigated  
 self-similar, spherically symmetric perfect fluid of models using the homothetic 
approach. They introduce dimensionless variables, so that the
number of equations in the coupled system of autonomous differential equations
is reduced, with the resulting reduced phase-space being compact and regular.
In this way the similarities with the equations governing hypersurface
orthogonal models, and in particular spatially homogeneous models (WE,
Nilsson and Ugglas 1997), can be exploited. In their approach,
all equilibrium points are hyperbolic, in contrast to the earlier work in
which Bogoyavlenski used non-compact variables (which resulted in parts
of phase-space being ``crushed'').  The spatially self-similar case 
was studied by Goliath et al. (1998a) and the timelike case, which contains the
more interesting physics (e.g., shocks and sound-waves), was studied
by Goliath et al. (1998b). 

The Schwarzschild approach was adopted by Ori and Piran (1990) and more
recently by Maison (1995). In order to obtain physically
reasonable models, spacetimes are often required to be asymptotically flat. 
Since asymptotically flat spacetimes are not self-similar, one therefore
needs to match a self-similar interior region to a non-self-similar
exterior solution. This is usually taken to be
Schwarzschild, in which case Schwarzschild coordinates are better
suited to finding global solutions. This approach is also suitable
for solving the equations of motion for (radial) null geodesics, enabling the
causal structure of spacetime to be studied. Consequently it was
used by Ori and Piran (1990) since one of their primary goals was to study
naked singularities and test the cosmic censorship hypothesis (Section~4A).
However, the Schwarzschild coordinates break down at $t=0$.   {\it Null}
coordinates can also be used to analyse the global structure (Henriksen and
Patel 1991) and these are often used in the study of spherically symmetric
scalar fields 
(Choptuik 1993, Gundlach 1997).

\subhead{B. General Features of Similarity Solutions}\endsubhead

Throughout the subsequent discussion we will use comoving coordinates, so
the metric in the spherically symmetric situation can be written in the
form
$$ds^{2}= -e^{2\nu}\,dt^{2}+e^{2\lambda}\,dr^{2}+R^{2}\,d\Omega^{2},\q
        d\Omega ^2 \equiv d\theta^{2}+\sin^{2}\theta \,d\phi^{2} 
\tag3.1$$
where $\nu$, $\lambda$ and $R$ are functions of $r$ and $t$. The equations
have a ``first integral'' $m(r,t)$ which can be interpreted as the mass within
comoving radius $r$ at time $t$. There is also a dimensionless quantity
E(r,t) which represents the total energy per unit mass for the shell with
comoving
coordinate r. Unless $p=0$, both these quantities decrease with increasing
$t$ because of the work done by the pressure.

Spherically symmetric homothetic solutions were first
investigated by Cahill and Taub (1971), who showed that by a suitable
coordinate transformation they can be put into a form in which all
dimensionless quantities such as $\nu$, $\lambda$, $E$ and
$$S\equiv\frac{R}{r}, \q M\equiv\frac{m}{R},\q
   P\equiv pR^{2},\q W\equiv\mu R^{2} \tag3.2$$
are functions only of the dimensionless similarity variable $z\equiv r/t$. 
The homothetic vector in these coordinates is 
$$
 \xi ^a \frac{\partial}{\partial x^a} = t\frac{\partial}{\partial t} +
r\frac{\partial}{\partial r}.
\tag3.3$$
Values of z
for which $M=1/2$ correspond to a black hole or cosmological apparent horizon
since the congruence of outgoing null geodesics have zero divergence.
Another important quantity is the function
$$ V(z)=e^{\lambda -\nu}z\;, \tag3.4 $$
which represents the velocity of the fluid relative to spheres of 
constant $z$. These spheres contract relative to the fluid for $z<0$ and
expand for $z>0$. The homothetic vector is timelike for $V<1$ and spacelike
for $V>1$. Special significance is attached to values of z for which
$|V|=\sqrt{\alpha}$ and $|V|=1$. The first corresponds to a sonic point
(where the pressure
and density gradients are not uniquely determined), the second to a Cauchy
horizon (either a black hole event horizon
or a cosmological particle horizon).

We have seen that the only barotropic equation of state compatible with the
similarity ansatz is
one of the form $p=\alpha \mu$. It is convenient to introduce a
dimensionless function $x(z)$ defined by
$$
x(z)\equiv (4\pi \mu r^{2})^{-\alpha/(1+\alpha)} .\tag3.5 $$
The conservation equations $T^{\mu \nu}_{\q;\nu}=0$ can then be 
integrated to give
$$e^{\nu}=\beta x z^{2\alpha/(1+\alpha)},\q
e^{-\lambda}=\gamma x^{-1/\alpha} S^{2} \tag3.6$$
where $\beta$ and $\gamma$ are integration constants. The remaining field 
equations reduce to a set of ODEs for $x$ and 
$S$ in terms of the similarity variable [see eqns (2.11) 
to (2.15) of Carr and Yahil (1990)]. These specify integral curves in the
three-dimensional $(x, S, \dot{S})$ space (where a dot denotes $zd/dz$).
For a given equation of state parameter $\alpha$, there is therefore a
two-parameter family of spherically symmetric similarity solutions.

In $(x, S, \dot{S})$ space the sonic condition $|V|= \sqrt{\alpha}$
specifies a two-dimensional surface. Where a curve intersects this surface,
the equations do not uniquely
determine $\dot{x}$ (since the coefficient of $\dot{x}$ disappears in one
of them), so there can be a number of different solutions passing
through the same point. However, only integral curves which pass through a
line  $Q$ on the sonic surface are ``regular'' in the sense that $\dot{x}$
is finite and they can be extended beyond there. The equations
permit just two values of $\dot{x}$ at each point of $Q$ and 
there will be two associated values of $\dot{V}$.  On some parts
of $Q$ these 
values will be complex (corresponding to a ``focal'' point), so the solution 
will still be unphysical. Otherwise both values of $\dot{V}$ will 
be real and at least one of them will 
be positive.  If both values of $\dot{V}$ are positive (corresponding to a
``nodal'' point), the smaller one is 
associated with a $1$-parameter family of solutions, while the larger one
is associated with an isolated solution.  If one of the values of $\dot{V}$ is
negative (corresponding to a ``saddle'' point), both values are associated
with isolated solutions. This behaviour has been analysed in detail by
Bogoyavlenski (1977), Bicknell and Henriksen (1978), Carr and Yahil (1990)
and Ori and Piran (1990).

On each side of the sonic point, $\dot{x}$ may have either of the two 
values.  If one chooses different values for $\dot{x}$, there will be a 
discontinuity in the pressure gradient.  If one chooses the same value,
there may still be a discontinuity in the higher derivatives of $x$.  Only
the isolated solution and a single member of the one-parameter family of
solutions  are analytic. This contrasts with the case of a shock, where $x$
is itself discontinuous  (Cahill and Taub 1971, Bogoyavlenski 1985, Anile
et al. 1987, Moschetti 1987).  One can show that the part of $Q$ containing
nodes, for which there is a one-parameter family
of solutions, corresponds to two ranges of values for $z$. One range 
$(z_1 < z < z_2)$ lies to the left of the Friedmann sonic point $z_F$ and
includes the static sonic point $z_S$. The other range $(z>z_3)$ includes
the Friedmann sonic point $z_F$. The values of $z_1$, $z_2$ and $z_3$
depend on
$\alpha$. The ranges for $\alpha =1/3$ are indicated in Figure (1); in this
case,  $z_2=z_S$ and $z_3 =z_F$.

Carr and Coley (1998a; CC) have classified the $p=\alpha \mu$ spherically
symmetric similarity solutions completely. The key steps in their analysis
are: (1) a complete analysis of the dust solutions, since this
provides a qualitative understanding of certain features of the general
solutions with pressure; (2) an elucidation of the link between the $z>0$
and $z<0$ solutions;
(3) a proof that, at large and small values of $|z|$, all similarity
solutions must have an asymptotic form in which $x$ and $S$ have a
power-law dependence on $z$; and (4) a demonstration that there are only
three such power-law solutions (apart from the 
trivial flat solution).

\subhead{C. Exact Power-Law  Similarity Solutions}\endsubhead

We first discuss the power-law models explicitly since they play a central
role in what follows. We will assume $z>0$ but the equations below can be
easily extended to the $z<0$ regime by replacing $z$ by $|z|$ and reversing
the sign of $V$.

$\bullet$ The $k=0$ Friedmann solution. For this one can choose $\beta$ and
$\gamma$
in equation (3.6) such that
$$x=z^{-2\alpha/(1+\alpha)},\q
        S=z^{-2/[3(1+\alpha)]} \tag3.7 $$
and then
$$
\mu = \frac{1}{4\pi t^2},\q
V=\left(\frac{1+3\alpha}{\sqrt{6}}\right)
z^{(1+3\alpha)/[3(1+\alpha)]}. \tag3.8$$
One can put the metric in a more familiar form by making a transformation
of the radial coordinate.  

$\bullet$ A self-similar Kantowski-Sachs (KS) model. For each $\alpha$
there is a unique self-similar KS solution and this can be put in the form
$$
S=S_* z^{-1}, \q x=x_* z^{-2\alpha/(1+\alpha)} \tag3.9$$
where $x_*$ and $S_*$ are constants determined by $\alpha$. One can take
$\beta$ and $\gamma$ to have the same values as in the Friedmann solution
for
$\alpha < 0$ and $i$ times those values for $\alpha > 0$. 
One then has
$$
\mu t^{2} =  \left( \frac{1}{3|\alpha|}\right)^{(1+\alpha)/(\alpha-1)},\q
   V = -\frac{(1-\alpha)(1+3\alpha)^2}{2 \sqrt{6}\alpha} \left(
\frac{1}{3|\alpha|}\right)^{-2 \alpha/(1-\alpha)} z^{(1+ 3
\alpha)/(1+\alpha)}
\tag3.10$$
and one can again put the metric in a familiar form with a radial
coordinate transformation. However, it should be stressed that
only solutions with $\alpha < -1/3$ are physical: for $0< \alpha <1$, $V$
is negative and this corresponds 
to tachyonic solutions (i.e., the t coordinate is spacelike and the r
coordinate is timelike); for $-1/3<\alpha<0$, $V$ is positive but $S$ is
imaginary. The mass is also negative for $-1/3<\alpha<0$. For $\alpha <
-1/3$, the mass is positive and the coordinates play their usual roles.
Although such solutions have negative pressure and violate the strong
energy condition, they may be relevant in the early Universe.

$\bullet$ A self-similar static solution. In this case 
$$
x=x_o,\q S=S_o \tag3.11$$
where the constants $x_o$ and $S_o$ are determined by $\alpha$, so there is
just one static solution for each equation of state. The other interesting
functions are  
$$\mu = x_o^{-(1+\alpha)/\alpha}r^{-2},\q
   V = x_o^{-(1-\alpha)/2\alpha}z^{(1- \alpha)/(1+\alpha)}. \tag3.12
$$
The metric can be put in an explicitly static form under an appropriate
change of variables. Misner and Zapolsky (1964) 
first found this solution (also see Oppenheimer and Volkoff 1939) but did not appreciate its self-similarity; it
has subsequently been studied by Henriksen and Wesson (1978) and Carr and
Yahil (1990).  Note that there is a naked singularity at the origin.

There is an interesting connection between the static and KS
solutions:  if one interchanges the $r$ and $t$ coordinates in the static
metric and also changes the equation of state parameter to
$$\alpha' = -\frac{\alpha}{1+2\alpha}, \tag3.13$$
one obtains the KS metric. For a static solution with a normal equation of
state $(1>\alpha>0)$, $\alpha'$ must lie in the range $-1/3$ to $0$, so
negative pressure (negative mass) KS solutions can also be interpreted as
positive pressure (positive mass) static solutions.

The forms of $V(z)$ for the Friedmann solution, KS and static solutions in
the $\alpha =1/3$ case is shown in Figure (1). The full family of
similarity models comprises solutions asymptotic to these at large and
small $|z|$. To study their asymptotic behaviour, one introduces functions
$A(z)$ and $B(z)$ defined by
$$x \equiv x_i e^A,\q
   S\equiv x_i e^B, \tag3.14$$
where $x_i$ is given by equation (3.7) in the Friedmann case, equation
(3.9) in the KS case and equation (3.11) in the static case. The ODEs for
$x$ and $S$ then become ODEs for $A$ and $B$.
The solutions in each family can be specified by the values of $A$ and $B$
as $|z| \rightarrow \infty$ (denoted by $A_\infty$ and $B_\infty$) and
their 
values as $|z| \rightarrow 0$ (denoted by $A_0$ and $B_0$), although these
values may not be independent. The form of the 
full family of solutions in the $\alpha = 1/3$ case is summarized in Figure
(2).

\subhead{D. Asymptotically Friedmann Solutions}\endsubhead 

In the supersonic (large $z$) regime, asymptotically Friedmann similarity
solutions are described by the single parameter $B_\infty$ since
$A_\infty =0$. The solutions are overdense relative to the Friedmann
solution for $B_\infty < 0$ and underdense for $B_\infty > 0$. If
$B_\infty$ is sufficiently negative, $V$ reaches a minimum value and then
rises again to infinity as z decreases. Such solutions contain black holes
and were originally studied because there was
interest in whether black holes could grow at the same rate as the particle
horizon. Carr and Hawking (1974) showed that such solutions
exist for radiation ($\alpha=1/3$) and dust ($\alpha=0$) 
but only if the universe is asymptotically rather than exactly 
Friedmann ($B_\infty \neq 0$); i.e., black holes formed through purely
local processes
in the early Universe cannot grow as fast as the particle horizon.
Carr (1976) and Bicknell and Henriksen (1978a) then extended this result to
a general $0<\alpha<1$ fluid. Lin et al. (1976) claimed that there {\it is}
a
similarity solution in an exact Friedmann universe for the special case of
a stiff fluid ($\alpha=1$) but Bicknell and Henriksen (1978b) showed that
this requires the inflowing material to turn into a null fluid at the event
horizon. In fact, 
for fixed $\alpha$, it is now known that {\it all} subsonic solutions which
can be attached to an exact Friedmann model via a sound wave are
non-physical:
as one goes inward from the sound-wave they either 
enter a negative mass regime or reach another sonic point at which
the pressure diverges (Bicknell and Henriksen 1978a). 

It is likely that 
asymptotically Friedmann solutions which contain black holes are supersonic
everywhere (in the sense that V never falls below $1/\sqrt{\alpha}$),
although this has not been rigorously proved.  However, all the solutions
with $B_\infty$ exceeding some critical negative value $B_\infty^{crit}$
reach the sonic surface and those which do so with $z_1 < z < z_2$ or $z >
z_3$ may be attached to the origin by subsonic solutions. The latter are
also described by a single
parameter and this can be taken to be $A_0$, solutions being overdense
relative to the Friedmann solution for $A_0 < 0$ and underdense for $A_0 >
0$.
These transonic solutions represent density fluctuations in a flat
Friedmann model which
grow at the same rate as the particle horizon (Carr and Yahil 1990). While
there is a continuum of regular underdense solutions, regular overdense
solutions only occur in successive and very narrow bands (with just one
solution per band being analytic at the sonic point). The overdense
solutions also exhibit oscillations in
the subsonic region, with the number of oscillations identifying the band.
The existence of these bands was first pointed out by Bogovalenski (1985)
and also studied by Ori and Piran (1990). The  band structure arises even
in the Newtonian situation (Whitworth and Summers 1985). The higher bands
are all nearly static near the sonic point but they
deviate from the static solution as one goes towards the origin. 

The forms of S(z) and V(z)
in the general $\alpha$ asymptotically Friedmann solutions are indicated in
Figure (3). The curves are here parametrized by the ``asymptotic energy''
parameter ($E_\infty$), which is related to $B_\infty$ by
$$
E_\infty = \frac{1}{2} (e^{6B_\infty} -1). \tag3.15$$
This is a more convenient parameter if one wishes to relate the asymptotically
Friedmann solutions to the other ones discussed below. The $z>0$ solutions
correspond to models which start from an initial Big Bang singularity at
$z=\infty$ ($t=0$) and then either expand to infinity as $z \rightarrow 0$
($t \rightarrow \infty$) for $E_\infty>E_{crit}$ or recollapse to a black
hole at some positive
value of z for $E_\infty<E_{crit}$. Here $E_{crit}$ is related to
$B_\infty^{crit}$ by equation (3.15). In the dust case, $E_{crit}=0$
because there is no pressure to stop the collapse if the energy is
negative. However, if there is pressure, $E_{crit}<0$ and we have
seen that there are overdense solutions without black holes 
which expand to infinity providing $E_\infty$ lies in narrow bands between
$E_{crit}$ and $0$; outside these bands the solutions
are irregular at the sonic point. The underdense solutions (with
$E_\infty>0$) are all regular at the sonic point. The $E_\infty<E_{crit}$
solutions contain a black hole event horizon and a cosmological particle
horizon for values of $E_\infty$ exceeding another critical value $E_*$
(corresponding to $V_{min}=1$). Note that the mass function M has a minimum
below 1/2, whatever the value of $E_\infty$, so there is always a black
hole and cosmological apparent horizon.   

The analysis is trivially extended to the $z<0$ regime since the solutions
are symmetric in z, as illustrated in Figure (3). Since $r$ is always 
taken to be positive, the $z<0$ solutions are
the time-reverse of the $z>0$ ones. Thus the $E_\infty>E_{crit}$ models
collapse from an infinitely dispersed initial state to a big crunch
singularity as z decreases from 0 to $-\infty$ (i.e. as t increases from
$-\infty$ to 0), while the $E_\infty<E_{crit}$ models emerge from a white
hole and are never infinitely dispersed.

\subhead{E. Asymptotically Kantowski-Sachs Solutions}\endsubhead

In this case, the asymptotic behaviour depends on the value of 
$\alpha$. For $0< \alpha < 1$, the solutions can be characterized by a single
parameter and this can be taken to be $A_\infty$ in the supersonic regime
and $A_0$ in the subsonic regime. However, 
there are only isolated solutions at a sonic point, so solutions which hit
the sonic surface are unlikely to be regular there. In any case, these
solutions - like the KS similarity solution itself - are presumably
unphysical. For $-1< \alpha < - 1/3$, there is no sonic point and the
solutions can be characterized by either $A_0$ or $A_\infty$. Note that the
asymptotic energy in all solutions is $E_\infty = -1/2$.
 
The asymptotically KS models were studied by Carr and Koutras (1993), who
integrated the equations numerically for $\alpha = 1/3$ and $\alpha = -1/2$. Figure (2) shows the form of $V(z)$ in the former case; although the
meaning of these solutions is unclear (since they are unphysical), they are
still of
mathematical interest in that they serve to fill in the $V(z)$ solution space.  
The solutions with $-1<\alpha < -1/3$ may be relevant in the early Universe
due to inflation or particle production effects. In
particular, they may be related to the growth of $p>0$ bubbles formed at a
phase transition in a $p<0$ cosmological background (Wesson 1986). For
example, Henriksen et al. (1983) have shown 
that a bubble in a 
de Sitter background can be modelled by a KS solution (although this
involves similarity of the second kind since the de Sitter model
contains a scale--see Section 5B).
Generalized negative-pressure KS similarity solutions (in which the
equation of state is different from $p=\alpha \mu$ and which can be
interpreted as a mixture of false vacuum and dust) have been studied by
Ponce de Leon (1988) and Wesson (1989). 

A rather peculiar feature of the asymptotically KS solutions is that the mass
can go negative. Indeed, this is a general feature of similarity solutions
and can occur even for $V>0$ (e.g., when one has shell-crossing). This may
seem unphysical but - in the context of the big bang model - Miller (1976)
has given a possible interpretation in terms of ``lagging'' cores. She
gives an explicit example of an $\alpha =1$ self-similar solution for which
the mass goes negative. In the $\alpha =1/3$ case (but only in this case), one
can show that there is a 
well-defined curve in the $V(z)$ diagrams where $M=0$ and this is shown in
Figure (1). This curve has two parts: the upper part (with $V > 0$) is
relevant for asymptotically Friedmann solutions, while the lower part (with
$V < 0$) is 
relevant for asymptotically KS solutions. $M$ is negative in between the
two parts and this region includes the KS solution itself (as expected).

\subhead{F. Asymptotically Static Solutions}\endsubhead

For $0<\alpha<1$, the asymptotically static solutions are described by two
parameters at large values of $z$. These can be taken to be $A_{\infty}$
and $B_{\infty}$, which in this case can be chosen independently. This also
determines the 
asymptotic energy parameter
$$
E_\infty = \frac{(1+\alpha)^2}{2(1+6\alpha + \alpha^2)}e^{6B_\infty -
2A_\infty /\alpha} - \frac{1}{2}. \tag3.16$$
Such solutions are of particular interest because they represent the most
general asymptotic behaviour.  As discussed in Sections 4A and 4B, they are
also associated with the formation of naked singularities and the occurence
of critical phenomena in gravitational collapse. However,
describing these solutions as ``asymptotically static'' at large $z$ is
rather misleading because one can show that the velocity of the fluid
relative to the constant R surfaces, 
$$
V_R = \frac{V \dot{S}}{S+\dot{S}}\;\; , \tag3.17$$
is generally non-zero as $z \rightarrow \infty$. Indeed, the 
asymptotic value of $V_R$ can also be expressed in terms of $A_{\infty}$
and $B_{\infty}$, and this is zero only for a one-parameter subfamily of
solutions. This agrees with the description of Foglizzo and Henriksen
(1993), who term such solutions ``symmetric''. CC describe the more general
solutions as asymptotically ``quasi-static'' since they still have
$\dot{S}$ and $\dot{x}$ tending to 0 at infinity. There are no
asymptotically quasi-static solutions at
the origin since all solutions must be either exactly static,
asymptotically Friedmann or asymptotically KS at small $z$.

A proper understanding of these solutions requires that one allows
for both positive and negative values of $z$. This is because the solutions
necessarily span both regimes, as illustrated by the form of $S(z)$ in
Figure (4). This shows that the introduction of the second parameter (which
CC term D) has two important consequences. First, although one still has
expanding and collapsing solutions for $D>0$, the Big Bang singularity
occurs at $z=-1/D$ rather than at $z=\infty$ (i.e., before $t=0$), while
the Big Crunch singularity occurs at $z=+1/D$ (i.e., after $t=0$). Second,
the solution is asymptotic to a quasi-static model rather than the
Friedmann model as $|z| \rightarrow \infty$. On the other hand, the second
parameter has relatively little effect in the subsonic regime, so one can
still use the results of the asymptotically Friedmann analysis here (at
least qualitatively). In particular, the model can only collapse from
infinity if $E_\infty$ is positive or 
lies in discrete bands if negative. Otherwise it must expand from an
initial white hole singularity at some negative value of z (just as in the
collapsing asymptotically Friedmann models) before collapsing to the black
hole singularity at $z=1/D$. The expanding solutions are just the time
reverse of the collapsing solutions. 

The form of V(z) in the $D>0$ collapsing solutions is also illustrated in
Figure (3). The solutions start with $V=0$ at $z=0$ and then, as z
decreases, pass through a Cauchy horizon (where $V=-1$) and then a sonic
point (where $V=-\sqrt{\alpha}$) before tending to the quasi-static form at
$z=-\infty$. They then jump to $z=+\infty$ and enter the $z>0$ regime. As z
further decreases, V first reaches a minimum and then diverges to infinity
when it encounters the Big Crunch singularity at $z_S=1/D$. The minimum of
V may be either above or below 1, depending on the values of $E_\infty$ and D, but
(as discussed in Section 4B) one necessarily has a naked singularity in the
latter case. The singularity forms with zero mass at $t=0$, but its mass
$m_S =(MSz)_S\;t$ then grows as t .
As pointed out by Ori and Piran (1990), such solutions have
an analogue in Newtonian theory (Larson 1969, Penston 1969). If the minumum
of V is below $\sqrt{\alpha}$, there is probably no solution since it would
need to pass through two further sonic points. At a given value of t, the
form of $\mu t^2$ specifies the density profile. If $z<0$, this corresponds
to an isothermal static distribution ($\mu \propto r^{-2}$) for $|z|>>1/D$
with a uniform core for $|z|<<1/D$. If $z>0$, it again corresponds to an
``isothermal'' static distribution for $z>>1/D$ but with a density
singularity at the origin.

Some of the two-parameter family of similarity solutions with pressure
have been studied numerically by Foglizzo and Henriksen (1993),
although they only focus on the collapsing ones. They confirm that the
solutions are described by two parameters at large $|z|$
and by one parameter at small $|z|$. They also identify the expected behaviour
at the sonic point. In their phase-space analysis, the orbits corresponding
to the overdense solutions converge on and then spiral around the static
solution for a while before heading to the origin. This corresponds to the
oscillations found by Carr and Yahil (1990) and Ori and Piran (1990).

\heading{4. Applications of Spherically Symmetric Similarity
Solutions}\endheading

\subhead{A. Gravitational Collapse and Naked Singularities}\endsubhead 

One of the major goals of classical general relativity in
recent years has been the study and testing of the
cosmic censorship hypothesis. This asserts, in very general
terms, that singularities which develop from regular initial conditions
have no causal influence on spacetime (Penrose 1969, Israel 1984).  
Until recently
most possible counter-examples to the cosmic censorship hypothesis 
have been restricted to spherically symmetric spacetimes 
which involve shell-crossing or shell-focussing, 
such solutions being globally naked for a suitable choice of initial data
(Eardley and Smarr 1979, Lake 1992).

Self-similarity is very relevant to this issue because most of the known
examples of shell-focussing
singularities involve exact homothetic solutions 
(Eardley et al. 1986, Zannias 1991, Lake 1992).
Indeed, it has been shown that a large subclass of self-similar solutions
have a central
singularity from which null geodesics emerge to infinity (Henriksen and
Patel 1991) and it has been argued that one might generally expect a naked 
singularity to have a horizon structure similar to that of the 
global homothetic solution (Lake 1992). The occurrence of naked
singularities in spherically symmetric, perfect fluid, self-similar
collapse has been studied by
Ori and Piran (1987, 1990), Waugh and Lake (1988, 1989), 
Lake and Zannias (1990), Henriksen and Patel (1991), and Foglizzo and
Henriksen (1993) [see also Ref. 2 in Lake 1992]. Their occurrence in the
spherically symmetric collapse of a self-similar massless scalar field has
been studied by Brady (1995). 

Most of the early work focussed on dust solutions
of the Tolman-Bondi class, including in particular the analytical work of
Eardley and Smarr (1979) and Christodolou (1984). Ori and Piran (1990; OP)
extended this work by studying spherically symmetric homothetic with
pressure.
For reasonable equations of state it might be expected that
pressure gradients would prevent the formation of shell-crossing
singularities (the situation is less clear for shell-focussing
singularities).  However, OP proved
the existence of a ``significant'' class of perfect fluid
self-similar solutions with a globally-naked central singularity. They
explicitly studied the causal nature of these solutions
by analysing the equations of motion for the radial null geodesics,
thereby demonstrating that the null geodesics emerge to 
infinity.  OP noted that these perfect fluid (non-dust) solutions 
might constitute the strongest known
counter-example to cosmic censorship, although they have not been shown to
be stable and consequently may not contradict the formulation of cosmic
censorship due to Penrose (1969).  Clearly it is important to study
the stability of perfect fluid spherically symmetric homothetic solutions
with naked singularities
with respect to non-self-similar perturbations (and eventually nonspherical
perturbations). We discuss this further in Section 4D.

Foglizzo and Henriksen (1993; FH) extend OP's analysis of the gravitational
collapse of homothetic
perfect fluid gas spheres with $p = \alpha \mu$ for all $\alpha$ between
0 and 1, partially 
utilizing the powerful dynamical systems approach of Bogoyavlenski
(1985). They show 
that the set of globally {\it analytic} naked solutions is discrete
but finite (and even empty for
large values of $\alpha$) and they confirm that the number of oscillations
in the flow is a good index, with the approach to the ``static'' solution
being recovered as this index grows. FH discuss how
the initial part of the ``precursor'' singularity (OP, Lake 1992), which is 
the only component which can become naked, is  
formed from initially inwardly
directed trajectories. 

Recent work, both theoretical
(Barrabas et al. 1991) and numerical (Shapiro and Teukolsky 1992),
has shown that naked singularities 
may also arise in non-spherical collapse.
In the latter work, the axisymmetric collapse of a prolate
cluster of collisionless matter to a singular ``spindle state''
was studied. However, these solutions are not self-similar.
FH also considered planar homothetic collapse and found that, in this case, the 
singularity is never naked.

\subhead{B. Critical Phenomena}\endsubhead

One of the most exciting developments in general relativity in recent
years has been the discovery of critical phenomena in gravitational
collapse. This first arose in studying the gravitational collapse of a
spherically symmetric massless (minimally coupled) scalar field
(Choptuik 1993). If one considers a family of imploding scalar wave
packets whose strength is characterized by a continuous parameter $l$,
one finds that the final outcome is either gravitational collapse for
$l > l^*$ or dispersal, leaving behind a regular spacetime, for $l <
l^*$. For $(l-l^*)/l^*$ positive and small, the final black hole mass
obeys a scaling law
$$
M_{BH} = C(l-l^*)^{\eta} \tag4.1$$
where C is a family-dependent parameter and $\eta=0.37$ is
family-independent. Initial data with $l = l^*$ leads to a critical
solution which exhibits ``echoing''. This is a {\it discrete}
self-similarity (DSS) in which all dimensionless variables $\Psi$ repeat
themselves on ever-decreasing spacetime scales:  $\Psi(t,r) = \Psi(e^{-n
\Delta} t, e^{-n \Delta} r)$ where n is a positive integer and
$\Delta=3.44$. Near-critical initial data first evolves towards the
critical solution, showing some echoing on small space scales, but then
rapidly evolve away from it to either form a black hole ($l > l^*$) or
disperse ($l < l^*$). 

The structure of the critical solution has been studied by Gundlach
(1995). He claims that the solution is unique providing the metric is
regular at the origin ($r=0$) and analytic across the past null-cone of
$r=t=0$. (The null-cone is also a sonic surface since the speed of
sound is the speed of light for a scalar field.) The solution has a
naked singularity at $r=t=0$. Gundlach also shows that the critical
solution is unstable: spherically symmetric perturbations about it
contain a single growing mode. A similar picture has emerged from the
numerical analysis of spherically symmetric field collapse for a
non-minimally coupled scalar field and for a self-interacting scalar
field (Choptuik 1994).  It also applies in the case of vacuum,
axisymmetric gravitational wave collapse (Abraham and Evans 1993).

Obtaining analytical solutions with a DSS is
difficult, so attempts have been made to elucidate critical phenomena
by studying spherically symmetric solutions which possess {\it continuous}
self-similarity (CSS; e.g., admit a homothetic vector). There is evidence
that CSS is a good approximation in the
near-critical regime, so mathematical simplification is not the only
motivation for these studies.  Spherically symmetric homothetic
spacetimes containing radiation (i.e., $\alpha = 1/3$) have been
investigated by Evans and Coleman (1994). They study models containing
ingoing Gaussian wave packets of radiation numerically and find
analogous non-linear behaviour to the scalar case. The scaling law even
has the same exponent $\eta$, although this appears to be a
coincidence since the exponent is different for other equations of
state (Maison 1995, Koike et al. 1995). They also obtain a homothetic critical solution which qualitatively
resembles the scalar DSS critical solution; in
particular, it has a curvature singularity at $t = r = 0$ but is regular
at  $r = 0$ and the
sonic point. As in the
scalar case, they find that the critical solution is an intermediate
attractor: as the critical point is approached, the evolution of the
fluid and gravitational field develops a self-similar region (given by
the exact critical solution) near the centre of collapse. However, only
a precisely critical model is described by this solution everywhere. Although
Evans and Coleman claim that their solution possesses a similarity of the
second kind, Carr and Henriksen (1998) show that the solution is actually 
of
the first kind.

Maison (1995) and Koike et al. (1995) have extended Evans and Coleman's
work to the more general
$p = \alpha \mu$ case. By considering spherically symmetric
non-self-similar perturbations to the critical solution, Maison also manages to
obtain the scaling behaviour indicated by equation (4.1) analytically. 
This analysis cannot be 
applied directly to the scalar field case, even though a  
massless scalar field can formally be described as a stiff ($\alpha
=1)$ perfect fluid whenever the gradient of 
the scalar field is timelike. However, spherically symmetric self-similar spacetimes
with a massless scalar field have been investigated by a number of
authors (Brady 1995, Koike et al. 1995, Hod and Piran 1997, Frolov
1997). The self-similar solution of Roberts (1989) is also relevant in
this context. This describes the implosion of scalar radiation from
past null infinity. The solution is described by a single parameter: it
collapses to a black hole when this parameter is positive, disperses to
future null infinity leaving behind Minkowski spacetime
when it is negative, and exhibits a null singularity when it is zero.
Although this is reminiscent of the usual critical behaviour, the critical
solution is not an intermediate attractor since nearby solutions do not
evolve towards it.

Continuous self-similarity also arises in the collapse of a complex
scalar field (Hirschmann and Eardley 1995a, 1995b) or the axion/dilation
field found in string theory (Hamad\`e et al. 1995).  However, in these cases,
the CSS solutions are unstable and the universal
critical attractor is DSS (Hamad\`e et
al. 1995).  Gundlach (1995, 1997) has proposed running detailed collapse
calculations for more complicated matter models, such as the continuous
one-parameter family of $p = \alpha \mu$ perfect fluid models.  In one
parameter region the critical solution might be discretely
self-similar, while in another it might be continuously self-similar.
Parameter values may even exist in which two equally strong attractors
could coexist, perhaps leading to new interesting non-linear
behaviour.  For example, Choptuik and Liebling (1996) have studied a massless
scalar field in Brans-Dicke gravity (which is equivalent to two scalar 
fields with a particular coupling in general
relativity) and observed a transition from continuous to discrete self-similarity
in the intermediate attractor as the Brans-Dicke parameter is varied and 
Choptuik et al. (1996) have investigated 
Einstein-Yang-Mills collapse and again found discrete self-similarity 
at the blackhole threshold as well 
as another region of parameter space where the intermediate attractor is the 
$n =1$ static Bartnik-McKinnon
solution. In addition, Brady et al. (1998) have studied massive scalar field
collapse.

It is clearly important to relate these studies of homothetic solutions to
the earlier ones described in Section 3 and to identify the critical
solutions among  the complete family described in Carr and Coley (1998a).
The overdense asymptotically Friedmann solutions already exhibit some of
the features of the critical solution in that they are nearly static inside
the sonic point and exhibit oscillations. They are also regular at the
origin and at the sonic point. However, they cannot be identified with the
critical solution itself since they do not contain a naked singularity at
the origin. Nor can the static solution itself be so identified since it
has a naked singularity at $r=0$ for all t, whereas the critical solution
only has a singularity at the origin for $t=0$. 

To identify the critical solution, one needs to consider the full
two-parameter family of spherically symmetric similarity solutions. This
has been discussed by Carr and Henriksen (1998), who argue that the
critical solution should be identified with the collapsing solution for
which $V_{min} =1$ since various global studies of naked singularities in
these solutions 
(see references in Section 4A) have shown that the condition $V_{min} =1$
heralds the appearance of a naked singularity at the space-time origin. 
We have seen that the family of analytic, homothetic solutions that contain
(initially massless) black holes and naked singularities at the space-time
origin is a {\it discrete} one parameter set. These solutions can be
characterized by
the number of oscillations they contain in the subsonic region $n_\alpha$.
As $n_\alpha$ increases, the minimum value $V_{min}$ obtained by V in the 
region $z>0$ (where the singularity lies) decreases. 
Thus there is always a first $n_\alpha$ for which $V_{min}<1$. This value,
$n_\alpha^*$ say, then labels the 
threshold for the formation of massless black holes and naked singularities. 
FH give $n_\alpha^*=1$ for $\alpha=1/16$, $4$ for $\alpha=1/3$ and $6$ for 
$\alpha=9/16$. They surmise that $n_\alpha^*\rightarrow\infty$ as $\alpha 
\rightarrow 1$ but do not prove this assertion. For sufficiently small
$\alpha$, $n_\alpha^*=0$ (OP). 

Unfortunately, there is a problem with this simple criterion. First, the
discreteness in the family 
of analytic solutions means that none of them is likely to have 
$V_{min}=1$ precisely. This suggests that the critical solution is
likely to be $C^1$ rather than analytic at the sonic point and
it is not clear from the Evans-Coleman paper whether this is the case. Second, all
the known critical solutions have a {\it single} oscillation in the
subsonic regime. It is possible that single-oscillation solutions
can have $V_{min}=1$ if one allows non-analyticity at the
sonic point but that would suggest that the critical
solution may not be unique. The precise identification of the 
critical solution therefore remains uncertain and is the subject 
of further work (Carr et al. 1998).

\subhead{C. Self-Similar Voids}\endsubhead

A few years ago measurements of the Hubble constant $H_0$
obtained through studying Cepheid variables in galaxies in the Virgo 
and Leo clusters gave values of around 80 $km\; s^{-1} Mpc^{-1}$ (Pierce et
al.\  1994, Freedman et al.\  1994). 
In the standard Big Bang model without a cosmological constant this made it
hard to reconcile the age of the Universe with the ages of  
globular clusters (at least 12 Gyr). More recent estimates 
of $H_0$ using Cepheids yield values closer to 70 (Freedman 1997), but there is still
an age problem. However, it must 
be stressed that these large values of $H_0$ are all obtained 
within the relatively local distance of 100 Mpc, which is much less than
the horizon size of order 10 Gpc. Observations based on the
Sunyaev-Zeldovich effect for clusters (Birkinshaw and Hughes 1994) and the
time delay in gravitational lensed quasars (Rhee 1991, Roberts et al.\
1991) at much 
larger distances give lower values for the Hubble constant, which would be
compatible with the ages of globular clusters.

Several people have pointed out that the apparent discrepancy 
between the local and distant values of the Hubble constant can be 
reconciled if we live in a region of the Universe for which the local density 
is considerably less than its global value (Moffat and Tatarski 1992 and
1994, Nakao et al. 1995, Shi et al. 1996, Maartens et al. 1997). This could
also explain why the local density parameter (e.g., the density inferred
from the analysis of Virgocentric flow) is less than the global value that
would be required by inflation. Such a region will be described as a
``void'' even though it is not
completely empty. This suggestion might not seem too radical since we already 
know that the Universe contains large-scale voids (Geller and Huchra 1989), 
as well as large-scale flows (Lauer and Postman 1994). However, to resolve
the age problem, we need the local void to extend to at least 100 Mpc (so
that it 
includes the Coma cluster, which is assumed to have negligible deviation 
from the Hubble flow in the Cepheid estimates of $H_{0}$) and this is much 
larger than the typical void.

In analysing this proposal, one needs to assume a particular model 
for the void. Since the similarity hypothesis (discussed
in Section 4D) suggests that cosmological density perturbations may inevitably
evolve to self-similar form, it is natural to model the voids by the sort
of underdense asymptotically Friedmann solutions discussed in Section 3C.
Indeed Newtonian studies already support this suggestion. These show that
voids evolve towards self-similar form at late times, with most of the
matter piling up onto a surrounding shell (Hoffman et al. 1983, Hausman et
al. 1983, Bertschinger 1985). This applies whether the void is produced by
a cosmic
explosion (Schwartz et al. 1975, Ikeuchi et al. 1983) or merely evolves
from the primordial density perturbations. Bertschinger (1985) has applied
this idea to explain giant cosmic voids but finds that 
one needs more than the linear fluctuations which arise in the standard
hierarchical clustering scenario. 

Although the studies cited above were all Newtonian, voids have also been 
studied in the relativistic context; for extensive references see Sato (1984)
and Krasinski (1997). Indeed, a 
relativistic treatment is obligatory for voids whose radius is
non-negligible compared to the particle horizon size. Many such treatments
use the Tolman-Bondi solution to model the void as an underdense sphere
embedded in an Einstein-de Sitter and then 
determine the ratio of the global and local Hubble parameters (Wu et al.
1995, Suto et al. 1995). However, if primordial fluctuations arise from an
inflationary phase, it is also natural to consider fluctuations which are
only asymptotically rather than exactly Friedmann. This has motivated  Carr
and Whinnett (1997) to model cosmic voids by the underdense self-similar
asymptotically Friedmann solutions discussed in Section 3C. Related 
solutions have also been discussed by Tomita (1995, 1997a, 1997b). 

We have seen that the precise form of such self-similar voids depends upon 
the equation of state. After the time of decoupling at around $10^{5}$ yrs, 
the Universe can be treated as pressureless ``dust'' ($\alpha=0$).
In this case, there is no sonic point and the 
underdense self-similar solutions can be analysed analytically as 
a special case of the Tolman-Bondi solutions. Carr and 
Whinnett express the various Hubble and density parameter profiles in terms
of the (negative) energy parameter $E_\infty$.
Although they find that the local values of these parameters may indeed
differ considerably from their global values, they also note that the
origin of the self-similar dust solution is non-regular in that the
circumference function is
non-zero in the limit $r\rightarrow 0$ unless $E_\infty=0$. Thus the coordinate
origin is an expanding two-sphere and the solution must be patched onto a
non-self-similar solution at the centre. This produces
anomalous behaviour in the $r$-dependence of the Hubble parameter, which is
in contradiction with the observational data. 

Although the self-similar dust solution is not a viable model for a void
in the real Universe, the similarity hypothesis is not really expected to
apply in the dust situation since  
it probably requires the effects of pressure. It is therefore 
more natural to assume that a void only tends to self-similar form in the
radiation-dominated  ($\alpha=1/3$) era before decoupling.
Since the special conditions required for self-similar evolution in 
the radiation era are incompatible with self-similar evolution after 
decoupling, this suggests that one should merely use the self-similar 
radiation solution to set the initial conditions for the non-self-similar 
Tolman-Bondi evolution in the dust era. More precisely, the similarity
solutions specify the forms of $R$, $m$ and $E$ as functions
of $r$ along the decoupling hypersurface and the Tolman-Bondi equations
then give the evolution of $R(t,r)$ for each shell of constant $r$. From
this one can calculate the various Hubble parameters at any given epoch. 

Carr and Whinnett find models that are in agreement with the observational
data and clearly show a variation of the Hubble parameter with distance.
However, these models have a drawback. The strength of the 
initial radiation perturbation is determined by the single parameter $A_{0}$,
which fixes both the density contrast and the size of the void relative to
the particle horizon. To obtain a void that is large enough to contain
the Coma cluster at the present epoch it is necessary to choose a value for
$A_{0}$ which implies that the density
contrast at decoupling exceeds the mean perturbations allowed by the data
from the COBE satellite. In addition,
at the current epoch, the void has a local density parameter which is much
lower than the observed value. One can select a smaller value of $A_{0}$ to
produce the required density contrast but, in this case, the void radius is
too small.

\subhead{D. The Similarity Hypothesis}\endsubhead 

The ``similarity hypothesis'' proposes that, in a variety of physical
situations, solutions may naturally evolve to self-similar form even if
they start out more complicated. We have already mentioned several examples
of this. 
For example, we noted in Section 1 that self-similar asymptotics
can be obtained from dimensional considerations in a wide range of contexts
in fluid dynamics (Barenblatt and Zeldovich 1972) and 
we saw in Section 2 that self-similar solutions  act 
as asymptotic states in the context of spatially homogeneous 
cosmological models.  We also
noted in Section
2C the conjecture of Hewitt and Wainwright (1990) that self-similar
solutions may describe the 
asymptotic or intermediate dynamical behaviour of 
orthogonally transitive $G_2$ models. It is even possible that the
hypothesis extends to self-similarity of the second kind. For example, the
``cosmic no-hair theorem'' asserts that all
cosmological models asymptote to the de Sitter solution in the
presence of a cosmological constant (cf. Wald 1993). de Sitter spacetime is not
homothetic but it is self-similar in that it admits a kinematic self-similarity
of the zeroth kind  
(Coley 1997a; see also Section 5A).

In this section, we will consider the plausibility of this hypothesis in the 
spherically symmetric context. There are a variety of astrophysical and
cosmological situations (both Newtonian and relativistic) in which
spherically symmetric solutions seem to evolve to self-similar form. For
example, an explosion in a homogeneous background produces
fluctuations which may be very complicated initially but which
tend to be described more and more closely by a spherically
symmetric similarity solution as time evolves (Sedov 1967). We saw in 
Section 4C that this
applies even if the explosion occurs in an expanding cosmological 
background and it may indeed be a general feature of voids, whatever their
origin. Overdense regions in the
hierarchically clustering scenario in Newtonian cosmology may also tend to self-similar
form (Quinn et al. 1986, Frenk et al. 1988) due to non-linear effects (Gunn and Gott 1972, Gunn 1977, Fillmore and
Goldreich 1984, Bertschinger and Watts 1988, Syer 
and White 1997). In the non-cosmological context, a gravitationally bound
cloud collapsing from an initially
uniform static configuration may also evolve to self-similar form 
(Penston 1969, Larson 1969). 
This is understood theoretically, at least in a Newtonian context, as
arising from such processes as virialization, shell-crossing and violent
relaxation (Lynden-Bell 1967).  In addition, it is known that in general 
relativity  {\it all} static, spherically symmetric perfect fluid
solutions with $p = \alpha \mu$ are asymptotically self-similar 
(Collins 1985, Goliath et al.\ 1998b).

These considerations led Carr (1993) to propose the ``cosmological
similarity hypothesis''. This states says that, under certain circumstances
(e.g., non-zero pressure, high non-linearity, shell-crossing, processes
analogous to virialization), cosmological solutions will naturally evolve
to a spherically symmetric self-similar form, whatever the nature of the
primordial fluctuations. We saw an application of this in Section 4C.
Another application involves
the studies of hierarchical cosmological models (Wesson 1979, 1981, 1982).
In principle, this proposal
is directly testable using numerical studies of spherically symmetric
perturbations of Friedmann models with pressure (cf. Frauendinier and
Schmidt 1993),
but this has not yet been done.
 
Presumably a necessary (but not sufficient) condition for the similarity
hypothesis to be valid is that spherically symmetric similarity solutions
(or at least some subset of them) be {\it stable} to non-self-similar
perturbations. As a first step to studying this, Carr and Coley (1998b)
have therefore investigated the stability of spherically symmetric
similarity solutions within the more general class of spherically symmetric
solutions. Following Cahill and Taub (1971), they express all functions in
terms of the similarity variable z=r/t and the radial coordinate r and
regard these as the independent variables rather than r and t. They also
assume that the perturbations in S and x (defined in Section 3A) can be
expressed in the form
$$
S(z,r) = S_o (z)[1+S_1 (r)] , \q  x(z,r) = x_o (z)[1+x_1 (r)]   \tag4.3$$       
where a subscript 0 indicates the form of the function in the exact
self-similar case and a subscript 1 indicates the fractional perturbation
in that function (taken to be small; i.e., $S_1<<1$ and $x_1<<1$). The
perturbation equations 
for $S_1(r)$ and $x_1(r)$ can then be expressed as second order
differential equations in $r$ and Carr and Coley test whether a particular
similarity solution is stable by examining, for example, whether the
perturbation terms grow or decay at large values of $r$. It might seem more
natural to examine whether the solution grows or decays with time. However,
the t evolution is entirely contained within the z evolution (viz.
$\partial f/\partial t = -t^{-1}\dot{f}$), so if one wrote the perturbation
equations in terms of t and z (instead of r and z), one would get exactly
equivalent results.

Carr and Coley (1998b) come to the following conclusions:

$\bullet$ The asymptotically Friedmann solutions are stable providing
$\alpha > -1/3$. This directly relates to the issue of whether density
perturbations naturally evolve to self-similar form. Of course, it does not
prove the validity of the similarity hypothesis since only {\it small}
perturbations of self-similar models are considered. Neither do they
consider non-spherical perturbations, for which the equations would be even
more complicated. However, it does seem to be a rather general property of
perturbations in an expanding Universe that they tend to sphericity at late
times. 

$\bullet$ As already shown by Ori and Piran (1988), transonic similarity
solutions which are not analytic at the sonic point are unstable. This
relates to what they term the ``kink'' instability: non-analytic solutions
either develop shocks or are driven towards analytic ones. This also
applies in the Newtonian context (Whitworth and Summers 1985).

$\bullet$ The asymptotically Kantowski-Sachs solutions are only stable for
$-1<\alpha<-1/3$, corresponding precisely to the range in which the
solutions are
physical. This may relate to the formation of bubbles in an inflationary
scenario and hence to the instability of the inflationary phase itself.

$\bullet$ The stability of the asymptotically quasi-static solutions is
still undetermined but is clearly of great physical interest. In
particular, the stability
of the ones containing naked singularities presumably bears upon the 
cosmic censorship hypothesis (discussed in Section 4A), while the
stability of the critical solutions must 
relate to the results of previous studies (discussed in Section 4B)  
which show that the critical solutions are   unstable to a single mode.

 \heading{5.  Generalized Self-Similarity}\endheading

Self-similarity in the broadest sense refers to the situation in 
which a system is not restricted to be invariant under the relevant group action but
merely appropriately rescaled.  The basic condition for a manifold vector field
$\pmb{\xi}$ to be a self-similar generator is that there exist constants $d_i$
such that, for each independent physical field $\Phi^i$,
$${\Cal L}_\xi \Phi^i = d_i \Phi^i. \tag5.1$$
For each $i$, $d_i$ is a constant which (formally) is the scalar product of the dimensionality covector 
of $\Phi^i$ with respect to the rescaling algebra and some rescaling algebra vector  (Carter and Henriksen 1991).

In the Newtonian case, the physical fields consist of $\mu$, $p$ and $\phi$ (the Newtonian gravitational potential), and
$\pmb{\xi}$ is the generator of a self-similarity with respect to
a three-dimensional rescaling algebra vector (time, length and mass).
Since $\mu$, $p$ and $\phi$ are scalar fields, ${\Cal L}_\xi$ denotes their
directional derivative.  The ordinary
continuity and dynamical
evolution equations are preserved by the action of this three-parameter
rescaling group.  Unfortunately, in general the self-similarity
will not survive when additional laws governing $p$ (the equation of 
state) and $\phi$ (Poisson's equation) complete the system of equations  (Carter
and Henriksen 1991). In order
for the system to remain invariant, 
further restrictions are imposed.  The effective invariance of
the Poisson equation restricts one to a two-dimensional subalgebra. Further
restrictions then arise from the imposition of an equation of state, except 
in the pressure-free case. For example, the 
special polytropic case $p = p_0 \mu^\gamma$ (where $\gamma$ is the polytropic
index) effects reduction to a one-parameter rescaling subalgebra, 
resulting in a specific ``self-similar'' index.  It is through
such an index that a formal basis can be provided
for classifying self-similarity as first class or,  
more generally, second class (Barenblatt and Zeldovich 1972).

These Newtonian ideas have been
adapted to the relativistic context by Carter and Henriksen (1989; CH).
Clearly, the characterizing equations must be generalized 
to tensorial counterparts of the covariant form
$${\Cal L}_\xi \Phi^i_A = d_i \Phi^i_A ,\tag5.2$$
where the fields $\Phi_A$ can be scalar (e.g., $\mu$), vectorial (e.g., $u_a$) or
tensorial (e.g., $g_{ab}$).  In particular, in general relativity
the gravitational potential $\phi$ is replaced by the metric
tensor $g_{ab}$ and an appropriate definition of ``geometrical''
self-similarity is necessary.  In the seminal work by
Cahill and Taub (1971), the simplest generalization was 
effected, whereby the metric itself satisfies an equation
of the form (5.2); in this case, $\pmb{\xi}$ is a homothetic vector and this
corresponds to Zeldovich's similarity of the first kind.

However, in general relativity it is not the energy-momentum tensor itself 
that must satisfy (5.2); rather each of the 
physical fields making up the energy-momentum tensor
must separately satisfy an equation of this form.
For a fluid characterized by a 
timelike congruence $u_a$, the energy-momentum tensor can be
uniquely decomposed with respect to $u_a$ (Ellis 1971),
each component
having a physical interpretation in terms of
the energy, pressure, heat flow and anisotropic stress as
measured by an observer comoving with the fluid,
and each separately satisfying an equation of the form (5.2).
In the same way, if the metric can be uniquely,
physically and covariantly decomposed, then the homothetic condition can
be replaced by the condition that each
component must satisfy (5.2). For a fluid, the metric can be uniquely decomposed, through the projection tensor
$$h_{ab} = g_{ab} + u_a u_b. \tag5.3$$
This represents the projection of the metric into
the $3$-spaces orthogonal to $u^a$ (i.e., into the 
rest frame of the comoving observers). If $u_a$ is irrotational,
these $3$-spaces are surface-forming, the decomposition
is global and $h_{ab}$ represents the intrinsic metric of
these $3$-spaces.  The projection tensor is the first fundamental form of the hypersurfaces orthogonal to $u^a$. It can be regarded as the
relativistic counterpart of the Newtonian metric tensor,
when the flow-independent $u_a$ is defined as the relativistic counterpart 
of the preferred  (irrotational) Newtonian time covector $-t_{,a}$ (CH).

\subhead{A. Kinematic Self-Similarity}\endsubhead

By using arguments similar to these and, more
importantly, comparing with 
self-similarity in a continuous Newtonian medium, CH
have introduced the covariant notion of {\it kinematic self-similarity} in the
context of relativistic fluid mechanics.  A kinematic
self-similarity vector $\pmb{\xi}$ satisfies the conditions
$${\Cal L}_\xi u_a = \overline{\alpha} u_a, \tag5.4$$
where $\overline{\alpha}$ is a constant and
$${\Cal L}_\xi h_{ab} = 2h_{ab}. \tag5.5$$
$\pmb{\xi}$ has been normalized so that the constant in (5.5)
has been set to unity.  Evidently, in the case $\overline{\alpha} =1$, 
it follows that $\pmb{\xi} $ is a HV (Cahill and Taub 1971), corresponding
to self-similarity of the first kind (Barenblatt and Zeldovich 1972).
CH then argue that the case $\overline{\alpha} \neq 1$
is the natural relativistic counterpart of self-similarity
of the more general second kind, while $(\ol{\alpha} =0)$ corresponds
to self-similarity of the zeroth kind.

The parameter $\ol{\alpha}$ is the constant  proportionality
factor between the rates of dilation of the length-scales and time-scales. When $\overline{\alpha} \neq 1$ (i.e., when $\pmb{\xi}$ is not a HV), the
relative rescaling of space and time under ${\pmb \xi}$ are not the same.
When $\overline{\alpha} = 0$, there is space
dilation without time amplification. The parameter $\overline{\alpha}$ defined by (5.4) is equivalent to the self-similar index which arises in the 
Newtonian case under the usual normalization ($u^a t_{, a} =1$).

All of the physical fields must satisfy equations of the form (5.2).
In the case of a perfect fluid, in addition to 
the ``kinematic'' self-similarity condition (5.4) and the
``geometric'' self-similarity condition (5.5), the physical energy
and pressure must therefore satisfy the conditions
$${\Cal L}_\xi \mu = a \mu, \enskip {\Cal L}_\xi p = bp , \tag5.6 $$
where $a$ and $b$ are constants.
In the exceptional
pressure-free case we have   
$${\Cal L}_\xi T_{ab} = (2 \ol{\alpha} +a) T_{ab}. \tag5.7 $$
If $p =p(\mu)$, then one necessarily has a polytropic equation of state,
$p = p_0 \mu^\gamma$, and equations
(5.6) imply that $b = a  \gamma$.
Kinematic self-similarities in 
perfect fluid spacetimes have been extensively studied 
in Coley (1997a).  In particular, a set of ``integrability''conditions
for the existence of a proper kinematic self-similarity in such spacetimes 
was derived; these integrability conditions constitute a
set of further constraints arising from the compatibility
of the EFEs and equations (5.4)--(5.6).

\subhead{B. Examples}\endsubhead 

In the {\it spherically symmetric} case, CH
have shown that there exist  comoving coordinates in which the
self-similar generator is 
$$\xi^a \frac{\partial}{\partial x^a} = (\overline{\alpha} t + \overline{\beta}) \frac{\partial}{\partial t} + r \frac{\partial}{\partial r}, \tag5.9$$
and the metric is given by (3.1), where
$\nu, \lambda$ and
$S=R/r$ again depend only on the self-similarity coordinate $z$.
The metric is manifestly of the same form as in Cahill and Taub (1971) and the
resulting governing differential equations do indeed reduce to a system of ODEs.

For self-similarity of the {\it first} kind
(the homothetic case), $\overline{\alpha} =1$, $\overline{\beta}$ can be rescaled to zero, and $z = r/t$ as usual.
In the less-studied {\it zeroth} case, $\overline{\alpha} =0$, 
$\overline{\beta}$ can be rescaled to unity and $z = re^{-t}$.  Examples
of this case are provided by the solution of Henriksen,
Emslie and Wesson (1983), in which a dimensional constant (and hence a fundamental scale)
is introduced via the cosmological constant, and the solution of
Alexander et al. (1989), which represents self-similar perturbations of
a de Sitter universe.  These solutions also 
relate to the Kantowski-Sachs models with $\alpha = -\frac{1}{3}$ studied
in Section 3C. 

In the general case with $\overline{\alpha} \ne 0$ or $1$ and $\overline{\beta}$ rescaled to zero,
corresponding to self-similarity of the {\it second} kind,
the self-similarity coordinate is given by
$$z = r (\overline{\alpha} t)^{-1/\overline{\alpha}}. \tag5.10$$
An important example of this 
is provided by a class of zero-pressure perfect fluid models
(i.e., dust models in which $u^a$ is geodesic)  first studied by Lynden-Bell and Lemos (1988) and described
in detail by Henriksen (1989) and CH.
This class of models has recently been generalized to 
the non-zero-pressure case as follows (Benoit and
 Coley 1998a; BC).  Using the similarity variables defined above and the
same notation as in Section 3, 
the metric for $S + \dot{S} \neq 0$ is given by
$$ds^2 = - dt^2 + (S + \dot{S})^2 dr^2 + r^2S^2 d{\Omega}^2, \tag5.11$$
where $S(z)$ is a solution of the differential equation
$$   2\frac{\ddot{S}}{S} + \frac{\dot{S}^2}{S^2} + 2 \overline{\alpha} 
\frac{\dot{S}}{S} + k = 0, \tag5.12$$
and $k$ is a 
dimensional constant. This assumes that $\dot{S} \neq 0$ (i.e., the spacetime is not static) and that
$\overline{\alpha}$ is neither 0 nor 1 (i.e., $\pmb{\xi}$ is 
not a homothetic vector).  The comoving perfect fluid is described by
$$\mu = W(\xi) t^{-2}  , \quad p =  \frac{1}{4} k \overline{\alpha}^{-4} t^{-2}. \tag5.13$$
where $W(\xi)$ is a  function of $\pmb{\xi}$ given  
by BC. [This solution is similar to the spherically symmetric 
generalization of the Kantowski-Sachs model discussed by Wesson (1989) in which
$S + \dot{S} = 0$.] The dust solution of the Tolman family, described in CH, is obtained when $k=0$. We note that in the solution above there is a 
dimensional constant appearing in the pressure, a property that is 
characteristic of self-similarity of the second kind (Barenblatt and Zeldovich 1972). This class of perfect fluid solutions does not admit any homothetic vectors. It has been generalized
to the anisotropic fluid case by Benoit and Coley (1998b).

There is another case of potential interest in which the
parameter $\overline{\alpha}$ occurring in equation (5.4) approaches infinity,
and we could refer to this case as kinematic self-similarity of the
``infinite'' kind.  This case could be covariantly defined by equation (5.4)
with $\overline{\alpha}$ normalized and equation (5.5) with the 
right-hand side set to zero (i.e., ${\Cal L}_{\xi}h_{ab}=0$). Consequently $\pmb{\xi}$ represents a generalized ``rigid motion''. This case has
been investigated by Benoit et al. (1998).

Recently, kinematic self-similarity in perfect fluid spacetimes
has been studied in the spherically symmetric
case (Benoit and Coley 1998a), the 
plane symmetric and hyperbolic cases (Benoit et al. 1998) and the locally rotationally symmetric case (Sintes 1997).
Note that there is some evidence that  kinematic self-similar 
spherically symmetric solutions asymptote towards 
exact solutions that admit a HV. For example, if $p = \alpha \mu$, these models necessarily reduce to either
the exact flat Friedmann model or the static model, both of 
which admit a HV.  In general, kinematic self-similar 
models do not admit an equation of state.  However, {\it all } 
kinematic self-similar spherically symmetric models in which the energy density and the pressure separately 
satisfy the ``physical'' self-similar conditions (5.6) have been
shown to {\it asymptote to an exact solution that admits a HV} (BC). It is therefore possible that the exact Friedmann, 
Kantowski-Sachs and static homothetic solutions
play the same role in describing the asymptotic behaviour of generalized
self-similar solutions as they did for the homothetic models discussed in Section 3

Another generalization of homothety is so-called 
{\it partial homothety} (Tomita 1981; see also Tomita and Jantzen 1983)
and this corresponds to an {\it intrinsic} symmetry (Collins 1979).  Ponce de Leon (1993) has studied partial homotheties in spherically symmetric fluid
models and attempted to relate the existence of such a
symmetry to the notion of generalized self-similarity.
However,  Ponce de Leon's approach is not covariant: even in the spherically symmetric situation there are
ambiguities in the shear-free case.  In addition, the matter fields
do not satisfy equations like (5.2) [i.e., like (5.6)], so the models are not ``physically'' self-similar (Ponce de Leon 1988, 1989).

\heading{6. Final Remarks}\endheading

We conclude with some general remarks about outstanding problems and areas
of the subject which are likely to see exciting developments in the next
few years.

$\bullet$  We saw in Section 5B that the asymptotic analysis of the
spherically symmetric homothetic models discussed in Section 3 might be
applicable more generally. Most of this review has focussed on perfect
fluids with an equation of state of the 
form $p = \alpha\mu$, so our analysis does not cover more general 
self-similar perfect fluids (Anile et al. 1987) or anisotropic fluids
(Herrera and Ponce de L\'eon 1985a,b),
even though these may be of physical interest.  However, there are clearly
ways of extending the analysis. For example, it can be shown that a 
two-perfect-fluid model, in which the two (necessarily comoving) fluids
each
have an equation of state of the form
$p_i = \alpha_i \mu_i$ ($i = 1, 2$), is formally equivalent to a model with a
single perfect fluid that does not have an equation of state.  It is
plausible that perfect fluid models for which $p/\mu$ is asymptotically
constant have
the same asymptotic behaviour as the $p = \alpha \mu$ perfect fluid models
discussed here. This is indeed the case for the self-similar two-perfect-fluid model if
each of the two fluids separately satisfy the conservation equations (Carr
and Coley 1998a).  In addition, an anisotropic fluid solution in which
the fluid has the form of a perfect fluid asymptotically might also
display these properties. For example, a source consisting of a  
perfect fluid plus an electromagnetic field (satisfying the Einstein-Maxwell
equations) can be formally equivalent to an anisotropic fluid source. 

$\bullet$ It is clear that a full understanding of the relationship between
critical phenomena and self-similarity will yield important insights, even
though the precise relationship between continuous self-similar
solutions and critical phenomena remains controversial (Carr and Henriksen
1998, Carr et al. 1998). In particular, an existence proof for the
critical solutions, which so far have only been 
constructed numerically, may be possible (Brady 1995, Gundlach 1997).
It is still not clear why the critical solution is sometimes associated
with a discrete self-similarity and whether this type of similarity is more
generic than its continuous counterpart. Doubtless resolving these issues
will require a 
deeper understanding of discrete self-similarity in general.

$\bullet$  Much work remains to be done in understanding the status of the
similarity hypothesis, especially in the spherically symmetric context. In
particular, 
the conditions under which this hypothesis is likely to hold have still
not been identified.  On the other hand, this is a problem which is
ideally suited for both numerical studies   and empirical cosmological tests, so one can expect progress. Further
studies of the  stability of the self-similar solutions will likely yield
new insights.

\heading{Acknowledgments}  \endheading

We would like to thank J. Carot, M. Choptuik, M. Goliath, C. Gundlach, R.
Henriksen,  C. Hewitt, K. Lake, M. Mars, U. Nilsson, A. Sintes, C. Uggla,
J. Wainwright and the referee for helpful comments.
AAC would like to acknowledge
the Natural Sciences and Engineering Research Council for 
financial support.  BJC would like to thank the
Department of Mathematics, Statistics and Computing Science at
Dalhousie University for hospitality while this 
work was carried out.

\newpage

\heading{REFERENCES}\endheading

\roster
\item"" \hsk   A. M. Abrahams and C. R. Evans, 1993, Phys. Rev. Lett. {\bf 70}, 2980.

\item""\hsk D. Alexander, R. M. Green and A. G. Emslie, 1989, MNRAS,  {\bf 237}, 93.

\item"" \hsk A. M. Anile, G. Moschetti and O. I. Bogoyavlenski, 1987, J. Math. Phys. {\bf 28}, 2942.

\item""\hsk G. I. Barenblatt, 1952, Prikl. Mat. Mekh. {\bf 16}, 67.

\item""\hsk G. I. Barenblatt and Ya B. Zeldovich, 1972, Ann. Rev. Fluid 
Mech. {\bf 4}, 285.

\item"" \hsk J. D. Barrow and F. J. Tipler, 1986, {\it The Anthropic Principle} (Oxford University
 Press, Oxford).

\item""\hsk  C. Barrabes, W. Israel and P. S. Letelier, 1991, Phys. Lett. A {\bf 160}, 41.

\item""\hsk P. M. Benoit and A. A. Coley, 1998a, Class. Quantum Grav.

\item""\hsk P. M. Benoit and A. A. Coley, 1998b, J. Math. Phys.

\item""\hsk P. M. Benoit, A. A. Coley and A.M. Sintes, 1998, preprint.

\item""\hsk  E. Bertschinger, 1985, Ap. J. {\bf 268}, 17.

\item""\hsk E. Bertschinger and P. N. Watts, 1984, Ap. J. {\bf 328}, 23.

\item""\hsk G. V. Bicknell and R. N. Henriksen, 1978a, Ap. J. {\bf 219}, 1043.

\item""\hsk G. V. Bicknell and R. N. Henriksen, 1978b, Ap. J. {\bf 225}, 237.

\item""\hsk G. V. Bicknell and R. N. Henriksen, 1979, Ap. J. {\bf 232}, 670.

\item""\hsk M. Birkinshaw and J. P. Hughes, 1994, Ap. J. {\bf 420}, 33.

\item""\hsk O. I. Bogoyavlenski, 1977, Sov. Phys. JETP {\bf 46}, 634.

\item""\hsk O. I. Bogoyavlenski, 1985, {\it Methods in the Qualitative Theory of  Dynamical Systems in Astrophysics and Gas Dynamics} (Springer-Verlag).

\item""\hsk O. I. Bogoyavlenski and S. P. Novikov, 1973, Sov. Phys.-JETP
{\bf 37}, 747.

\item"" \hsk H. Bondi, 1947, MNRAS {\bf 107}, 410.

\item"" \hsk W. B. Bonnor, 1956, Z. Astrophys. {\bf 39}, 143.

\item""\hsk P. R. Brady, 1995, Phys. Rev. D {\bf 51}, 4168 .
 
\item""\hsk P. R. Brady, C. M. Chambers and S. M. C. V. Concalves, 1998, 
 Phys. Rev. D.

\item""\hsk M. Bruni, S. Matarrase and O. Pantano, 1995, Ap. J. {\bf 445}, 958.

\item""\hsk A. B. Burd and A. A. Coley, 1994, Class. Quantum Grav. {\bf 11}, 83.

\item""\hsk A. H. Cahill and M. E. Taub, 1971, Comm. Math. Phys. {\bf 21}, 1.

\item""\hsk J. Carot  and A. M. Sintes, 1997, Class Quantum Grav. {\bf 14}, 1183.

\item""\hsk J. Carot, J. da Costa and E. G. L. R. Vaz, 1994, J. Math. Phys. {\bf 35}, 4832.

\item""\hsk B. J. Carr, 1976,  Ph.D. thesis (Cambridge).

\item""\hsk B. J. Carr, 1993, preprint prepared for but omitted from {\it The Origin of Structure in the Universe}, ed. E. Gunzig and P. Nardone (Kluwer).

\item""\hsk B. J. Carr, 1997, in {\it Proceedings of the Seventh Canadian Conference on General Relativity and Relativistic Astrophysics},  ed. D. Hobill (Calgary Press).

\item""\hsk B. J. Carr and S. W. Hawking, 1974, MNRAS {\bf 168}, 399.

\item""\hsk B. J.  Carr and A. Yahil, 1990, Ap. J. {\bf 360}, 330.

\item""\hsk B. J. Carr and A. Koutras, 1992, Ap. J. {\bf 405}, 34.

\item""\hsk B. J. Carr and A. A. Coley, 1998a, Phys. Rev. D.

\item""\hsk B. J. Carr and A. A. Coley, 1998b, preprint.

\item""\hsk B. J. Carr and R. N. Henriksen, 1998, preprint.

\item""\hsk B. J.  Carr and A. Whinnett, 1997, MNRAS.

\item""\hsk B. J.  Carr, A. A. Coley, M. Goliath, U. S. Nilsson and C. Uggla, 1998, preprint.

\item""\hsk B. Carter and R. N. Henriksen, 1989, Ann. Physique Supp. {\bf 14}, 47.

\item""\hsk  B. Carter and R. N. Henriksen, 1991, J. Math. Phys. {\bf 32}, 2580.

\item""\hsk  W. Z. Chao, 1981, Gen. Rel. Grav. {\bf 13}, 625.

\item""\hsk  M. W. Choptuik, 1993, Phys. Rev. Lett. {\bf 70}, 9.

\item""\hsk  M. W. Choptuik, 1994, in {\it 
Deterministic Chaos in General Relativity}, ed. D. Hobill et al. 
(Plenum, New York).

\item""\hsk  M. W. Choptuik and S. Liebling, 1996, Phys. Rev. Lett. {\bf 77},
1424.

\item""\hsk  M. W. Choptuik, T. Chmaj and P. Bizon, 1996, Phys. Rev. Lett. {\bf 77}, 424.

\item"" \hsk D. Christodoulou, 1984, Commun. Math. Phys. {\bf 93}, 171.

\item""\hsk A. A. Coley, 1997a, Class. Quantum Grav. {\bf 14}, 87.

\item""\hsk A. A. Coley, 1997b, in {\it Proceedings of the Sixth Canadian Conference
on General Relativity and Relativistic Astrophysics}, eds. S. Braham, J. Gegenberg and R. McKellar, Fields Institute Communications Series (AMS), Volume 15, p.19 (Providence, RI). 

\item""\hsk A. A. Coley and B. O. J. Tupper, 1983, Ap. J. {\bf 271}, 1.

\item""\hsk A. A. Coley and B. O. J. Tupper, 1989, J. Math. Phys. {\bf 30}, 2616.

\item""\hsk A. A. Coley and B. O. J. Tupper, 1990, Class. Quantum Grav. {\bf 7}, 1961.

\item""\hsk A. A. Coley and J. Wainwright, 1992, Class. Quantum Grav. {\bf 9}, 651 
  
\item""\hsk A. A. Coley and R. J. van den Hoogen, 1994a, J. Math. Phys. {\bf 35}, 4117.

\item""\hsk A. A. Coley and R. J. van den Hoogen, 1994b, in {\it 
Deterministic Chaos in General Relativity}, ed. D. Hobill et al. 
(Plenum, New York).

\item""\hsk A. A. Coley and R. J. van den Hoogen, 1995, Class. Quantum Grav., to appear.

\item""\hsk A. A. Coley and J. Wainwright, 1998, preprint.

\item""\hsk A. A. Coley, R. J. van den Hoogen and R. Maartens, 1996, Phys. Rev. D. {\bf 54}, 1393.

\item""\hsk A. A. Coley, J. Iba\~nez and R.J. van den Hoogen, 1997, J. Math. Phys. {\bf 38}, 5256.  

\item""\hsk C. B. Collins, 1971, Comm. Math. Phys. {\bf 23}, 137.

\item""\hsk C. B. Collins, 1974, Comm. Math. Phys. {\bf 39}, 131.

\item""\hsk C. B. Collins, 1977, J. Math. Phys. {\bf 18}, 2116.

\item""\hsk C. B. Collins, 1979, Gen. Rel. Grav. {\bf 10}, 925.

\item""\hsk C. B. Collins, 1985, J. Math. Phys. {\bf 26}, 2268.

\item""\hsk C. B. Collins and J. M. Stewart, 1971, MNRAS {\bf 153}, 419.

\item""\hsk C. B. Collins and S. W. Hawking, 1973, Ap. J. {\bf 180}, 317.

\item""\hsk  C. B. Collins and J. M. Lang, 1987, Class. Quantum Grav. {\bf 4}, 61.

\item""\hsk L. Defrise-Carter, 1975, Comm. Math. Phys. {\bf 40}, 273.

\item""\hsk A. G. Doroshkevich, V. N. Lukash and I. D. Novikov, 1973, Sov. Phys.-JETP {\bf 37}, 739.

\item""\hsk C. C. Dyer, 1979, MNRAS {\bf 189}, 189.

\item""\hsk C. Eckart, 1940, Phys. Rev. {\bf 58}, 919.

\item""\hsk D. M. Eardley, 1974, Comm. Math. Phys. {\bf 37}, 287.

\item""\hsk D. M. Eardley and L. Smarr, 1979, Phys. Rev. D {\bf 19}, 2239.

\item""\hsk  D. M. Eardley, J. Isenberg, J. Marsden and V. Moncrief, 1986, Comm. Math. Phys. {\bf 106}, 137.

\item""\hsk G. F. R. Ellis, 1971, {\it Relativistic Cosmology}, in 
{\it General Relativity and Cosmology, XLVII Corso, Varenna, Italy} (1969), ed R. Sachs (Academic, New York).

\item""\hsk  G. F. R. Ellis and M. A. H. MacCallum, 1969, Comm. Math. Phys. {\bf 12}, 108.

\item""\hsk C. R. Evans and J. S. Coleman, 1994, Phys. Rev. Lett. {\bf 72}, 1782.

\item""\hsk A. Feinstein and J. Iba\~nez, 1993, Class. Quantum Grav. {\bf 10}, 93.

\item""\hsk J. A. Fillmore and P. Goldreich, 1984, Ap. J. {\bf 281}, 1.

\item""\hsk T. Foglizzo and R. N. Henriksen, 1993, Phys. Rev. D. {\bf 48}, 4645.

\item""\hsk J. Frauendiener and B. G. Schmidt, 1993, Gen. Rel. Grav. {\bf 25}, 373.

\item""\hsk W. L. Freedman et al., 1994, Nature {\bf 371}, 757.

\item""\hsk W. L. Freedman, 1997, preprint.

\item""\hsk C. S. Frenk et al., 1988, Ap. J. {\bf 351}, 10.

\item""\hsk A. V. Frolov, 1997, preprint.

\item""\hsk M. J. Geller and J. P. Huchra, 1989, Science {\bf 246}, 897.

\item""\hsk B. B. Godfrey, 1972, Gen. Rel. Grav. {\bf 3}, 3.

\item""\hsk M. Goliath, U. S. Nilsson and C. Uggla, 1998a, Class Quantum Grav. {\bf 15}, 167.

\item""\hsk M. Goliath, U. S. Nilsson and C. Uggla, 1998b, Class Quantum Grav.

\item""\hsk C. Gundlach, 1995, Phys. Rev. Lett. {\bf 75}, 3214.

\item""\hsk C. Gundlach, 1997, Phys. Rev. D. {\bf 55}, 695.

\item""\hsk J. E.  Gunn, 1977, Ap. J. {\bf 218}, 592.

\item""\hsk J. E. Gunn and J. R. Gott, 1972, Ap. J. {\bf 176}, 1.

\item""\hsk V. T. Gurovich, 1967, Sov. Phys. Doklady {\bf 11}, 569.

\item""\hsk G. Haager and M. Mars, 1998, Class. Quantum Grav.

\item""\hsk G.S. Hall and D. Steele, 1990, Gen. Rel. Grav. {\bf 22}, 457.

\item""\hsk G. S. Hall, D. J. Low and J. R. Pulham, 1994, J. Math. Phys.  {\bf 35}, 5930.

\item""\hsk R. S. Hamad\'e, J. H. Horne and J. M. Stewart, 1996, Class. Quantum Grav. {\bf 13}, 2241.

\item""\hsk M. A. Hausman et al., 1983, Ap. J. {\bf 270}, 351.

\item""\hsk R. N. Henriksen, 1989, MNRAS {\bf 240}, 917.

\item""\hsk R. N. Henriksen and P. S. Wesson, 1978a, Ap. Sp. Sci. {\bf 53}, 429.

\item""\hsk R. N. Henriksen and P. S. Wesson, 1978b, Ap. Sp. Sci. {\bf 53}, 445. 
\item""\hsk R. N. Henriksen, A. G. Emslie and P. S. Wesson, 1983, Phys. Rev. D {\bf 27}, 1219.

\item""\hsk R. N. Henriksen and K. Patel, 1991, Gen. Rel. Grav. {\bf 23}, 527.

\item"" \hsk L. Herrera and J. Ponce de L\'eon, 1985a, J. Math. Phys. {\bf 26}, 
2302.

\item"" \hsk L. Herrera and J. Ponce de L\'eon, 1985b, J. Math. Phys. {\bf 27}, 2987.

\item""\hsk C. G. Hewitt and J. Wainwright, 1990, Class. Quantum Grav. 
{\bf 7}, 2295.

\item""\hsk C. G. Hewitt and J. Wainwright, 1992, Phys. Rev. D {\bf 46}, 4242.

\item""\hsk C. G. Hewitt and J. Wainwright, 1993, Class. Quantum Grav. {\bf 10}, 99.

\item""\hsk C. G. Hewitt, J. Wainwright and S. W. Goode, 1988, Class. 
Quantum Grav. {\bf 5}, 1313.

\item""\hsk C. G. Hewitt, J. Wainwright and M. Glaum, 1991, Class. Quantum Grav. {\bf 8}, 1505.

\item""\hsk E. W. Hirschmann and D. M. Eardley, 1995a, Phys. Rev. D {\bf 51}, 4198.

\item""\hsk E. W. Hirschmann and D. M. Eardley, 1995b, Phys. Rev. D {\bf 52}, 
5850.

\item""\hsk S. Hod and T. Piran, 1997, Phys. Rev. D. {\bf 55}, R440.

\item""\hsk Y. Hoffman and J. Shaham, 1985, Ap. J. {\bf 297}, 16.

\item""\hsk R. J. van den Hoogen, 1995, Ph.D. Thesis (Dalhousie University).

\item""\hsk R. J. van den Hoogen, A. A. Coley and J. Iba\~nez, 1997, Phys. Rev. D. {\bf 55}, 1.

\item""\hsk L. Hsu and J. Wainwright, 1986, Class. Quantum Grav. {\bf 3}, 1105.

\item""\hsk  J. Iba\~nez, R. J. van den Hoogen and A. A. Coley, 1995, Phys. Rev. D {\bf 51}, 928.

\item""\hsk S. Ikeuchi, K. Tomisaka and J. P. Ostriker, 1983, Ap. J. {\bf 265}, 583.

\item""\hsk  W. Israel, 1984, Found. Phys. {\bf 14}, 1049.

\item""\hsk K. C. Jacobs, 1968, Ap. J. {\bf 153}, 661.

\item""\hsk R. T. Jantzen, 1984, {\it Cosmology of the Early Universe}, 
ed R. Ruffini and L. Fang (World Scientific, Singapore).

\item""\hsk R. T. Jantzen and K. Rosquist, 1986, Class. Quantum Grav. 
{\bf 3}, 281.

\item""\hsk  R. Kantowski and R. Sachs, 1966, J. Math. Phys. {\bf 7}, 443.

\item""\hsk T. Koike, T. Hara and  S. Adachi, 1995, Phys. Rev. Lett. {\bf 74}, 5170.

\item""\hsk A. Koutras, 1992, Ph.D. thesis (Queen Mary and Westfield College).

\item""\hsk D. Kramer, H. Stephani, M. A. H. MacCallum and E. Herlt, 
1980, {\it Exact Solutions of Einstein's Field Equations}
(Cambridge University Press, Cambridge).

\item""\hsk A. Krasinzki, 1997, {\it Physics in an Inhomogeneous Universe} (Cambridge
University Press, Cambridge).
 
\item""\hsk K. Lake, 1992, Phys. Rev. Lett. {\bf 68}, 3129.

\item""\hsk K. Lake and T. Zannias, 1990, Phys. Rev. D {\bf 41}, 3866.

\item""\hsk R. B. Larson, 1969, MNRAS {\bf 145}, 271.

\item""\hsk T. R. Lauer and M. Postman, 1994, Ap. J. {\bf 425}, 418.

\item""\hsk D. N. C. Lin, B. J. Carr and S. D. M. Fall, 1978, MNRAS {\bf 177}, 151.

\item""\hsk  X. Lin and R. M. Wald, 1989, Phys. Rev. D {\bf 40}, 3280.

\item""\hsk J. P. Luminet, 1978, Gen. Rel. Grav. {\bf 9}, 673.

\item""\hsk D. Lynden-Bell, 1967, MNRAS {\bf 136}, 101.

\item""\hsk  D. Lynden-Bell and J. P. S. Lemos, 1988, MNRAS {\bf 233}, 197.

\item""\hsk  P. K-H. Ma and J. Wainwright, 1994 in {\it Deterministic Chaos in
General Relativity}, ed. D. Hobill et al. (Plenum, New York).

\item""\hsk R. Maartens and S. D. Maharaj, 1986, Class. Quantum Grav. {\bf 3}, 1005.

\item""\hsk R. Maartens, N. P. Humphreys, D. R. Matravers and W. R. Stoeger,
1996, preprint.

\item""\hsk S. D. Maharaj, 1988, J. Math. Phys. {\bf 29}, 1443.

\item""\hsk D. Maison, 1996, Phs. Lett. B. 366, 82.

\item""\hsk C. B. G. McIntosh, 1975, Gen. Rel. Grav. {\bf 7}, 199.

\item""\hsk B. D. Miller, 1976, Ap. J. {\bf 208}, 275.

\item""\hsk C. W. Misner and H. S. Zapolsky, 1964, Phys. Rev. Lett. {\bf 12}, 635.

\item""\hsk J. W. Moffat and D. C. Tatarski, 1992, Phys. Rev. D. {\bf 45}, 3512.

\item""\hsk J. W. Moffat and D. C. Tatarski, 1995, Ap. J.. {\bf 353}, 17.

\item""\hsk G. Moschetti, 1987, Gen. Rel. Grav. {\bf 19}, 155.

\item""\hsk K. Nakao et al., 1995, Ap. J. {\bf 453}, 541.

\item"" \hsk U. S. Nilsson and C. Uggla, 1997, Class. Quantum Grav. {\bf 14}, 1965.

\item""\hsk A. Ori and T. Piran, 1987, Phys. Rev. Lett. {\bf 59}, 2137.

\item""\hsk A. Ori and T. Piran, 1988, Mon. Not. R. Astron. Soc. {\bf 234}, 821.

\item""\hsk A. Ori and T. Piran, 1990, Phys. Rev. D. {\bf 42}, 1068.

\item""\hsk  R. Penrose, 1969, Nuovo Cim. {\bf 1}, 252.

\item""\hsk M. V. Penston, 1969, MNRAS {\bf 144}, 449.

\item""\hsk M. J. Pierce et al., 1994, Nature {\bf 371}, 211.

\item""\hsk J. Ponce de Leon, 1988, J. Math. Phys. {\bf 29}, 2479.

\item""\hsk J. Ponce de Leon, 1990, J. Math. Phys. {\bf 31}, 371.

\item"" \hsk J. Ponce de Leon, 1991, MNRAS {\bf 250}, 69.

\item""\hsk  J. Ponce de Leon, 1993, Gen. Rel. Grav. {\bf 25}, 865.

\item""\hsk P. J. Quinn et al., 1986,  Nature {\bf 322}, 329.

\item"" \hsk G. F. R. N. Rhee, 1991, Nature {\bf 350}, 211.

\item""\hsk M. D. Roberts, 1989, Gen. Rel. Grav. {\bf 21}, 907.

\item"" \hsk D. H. Roberts et al., 1991, Nature {\bf 352}, 43.

\item"" \hsk K. Rosquist, 1984, Class. Quantum Grav. {\bf 1}, 81.

\item""\hsk K. Rosquist and R. T. Jantzen, 1988, Phys. Rep. {\bf 166}, 189.

\item""\hsk K. Rosquist, C. Uggla and R. T. Jantzen, 1990, Class. Quantum Grav. {\bf 7}, 625.

 \item""\hsk H. Sato, 1984, in {\it General Relativity and Gravitation}, p289, ed. 
B. Bertotti et al.\ (Reidel, Dordrecht).

\item"" \hsk J. Schwartz, J. P. Ostriker and A. Yahil, 1975, Ap. J. {\bf 202}, 1.

\item""\hsk L. I. Sedov, 1946, Prikl. Mat. Mekh. {\bf 10}, 241.

\item"" \hsk   L. I. Sedov, 1967, {\it Similarity and Dimensional Methods in Mechanics} (New York, Academic).

\item""\hsk S. L. Shapiro and S. A. Teukolsky, 1992, Phys. Rev. D {\bf 45}, 2006.

\item""\hsk X. Shi, L. M. Widrow and L. J. Dursi, 1996, MNRAS {\bf 281}, 565 (1996).

\item""\hsk I. S. Shikin, 1979, Gen. Rel. Grav. {\bf 11}, 433.

\item""\hsk S. T. C. Siklos, 1981, J. Phys. A {\bf 14}, 395.

\item"" \hsk A. M. Sintes, 1996, Ph.D thesis (University of the Balearic Islands).

\item"" \hsk A. M. Sintes, 1997, preprint, ``KSS LRS models''.

\item""\hsk Y. Suto et al., 1995, Prog. Theor. Phys. {\bf 93}, 839.

\item""\hsk Syer and White, 1997.

\item""\hsk A. H. Taub, 1972, {\it General Relativity, papers in honour of J. L. Synge}, ed. L.O'Raifeartaigh (Oxford University Press, London).

\item""\hsk A. H. Taub, 1973, Comm. Math. Phys. {\bf 29}, 79.

\item""\hsk G. I. Taylor, 1950, Proc. Roy. Soc. London {\bf A201}, 175.

\item""\hsk K. S. Thorne, 1967, Ap. J. {\bf 148}, 51.

\item"" \hsk R. C. Tolman, 1934, Proc. Nat. Acad. Sci. {\bf 20}, 169.

\item"" \hsk K.  Tomita, 1981, Prog. Theor. Phys. {\bf 66}, 2025.

\item""\hsk K. Tomita, 1995, Ap. J. {\bf 451}, 1.

\item""\hsk K. Tomita, 1997a, Phys. Rev. D. {\bf 56}, 3341.

\item""\hsk K. Tomita, 1997b, Gen. Rel. Grav. {\bf 29}, 815.

\item""\hsk K. Tomita and R.T. Jantzen, 1983, Prog. Theor. Phys. {\bf 70}, 886.

\item""\hsk C. Uggla, 1992, Class. Quantum Grav. {\bf 9}, 2287.

\item"" \hsk C. Uggla, R. T. Jantzen and K. Rosquist, 1995, Phys. Rev. D {\bf 51}, 5522.

\item""\hsk J. Wainwright, 1985, in {\it Galaxies, Axisymmetric Systems and Relativity}, ed. M. MacCallum (Cambridge University Press, Cambridge).

\item""\hsk J. Wainwright and L. Hsu, 1989, Class. Quantum Grav. {\bf 6}, 1409.
 
\item""\hsk  J. Wainwright and G. F. R. Ellis, 1997, {\it Dynamical systems in cosmology} (Cambridge University Press, Cambridge).
 
\item""\hsk J. Wainwright, W. C. W. Ince and B. J. Marshman, 1979, Gen. Rel. Grav. {\bf 10}, 259.

\item""\hsk J. Wainwright, A. A. Coley, G. F. R. Ellis and M. Hancock, 1998, 
Class. Quantum Grav. {\bf 15}, 331.

\item""\hsk R. M. Wald, 1983, Phys. Rev. D. {\bf 28}, 2118.

\item""\hsk B. Waugh and K. Lake, 1988, Phys. Rev. D {\bf 38}, 1315.

\item""\hsk B. Waugh and K. Lake, 1989, Phys. Rev. D {\bf 40}, 2137.

\item""\hsk P. S. Wesson, 1979, Ap. J. {\bf 228}, 647.

\item""\hsk P. S. Wesson, 1981, Phys. Rev. D {\bf 23}, 2137.

\item""\hsk P. S. Wesson, 1982, Ap. J. {\bf 259}, 20.

\item""\hsk P. S. Wesson, 1989, Ap. J. {\bf 336}, 58.

\item""\hsk A. Whinnett, B. J. Carr and A. A. Coley (1998). In preparation.

\item""\hsk A. Whitworth  and D. Summers, 1985, MNRAS, {\bf 214}, 1.

\item""\hsk X. Wu et al., 1995, preprint.

\item""\hsk K. Yano, 1955, {\it The Theory of Lie Derivatives} (North-Holland, Amsterdam).

\item""\hsk T. Zannias, 1991, Phys. Rev. D {\bf 44}, 2397.

\item""\hsk Ya. B. Zeldovich and A. S. Kompaneets, 1950, {\it Collection Dedicated to Joffe} {\bf 61}, ed. P.I. Lukirsky  (Izd. Akad. 
Nauk SSSR, Moscow).

\item""\hsk Ya. B. Zeldovich and Yu. P. Raizer, 1963, {\it Physics of Shock Waves and
 High Temperature Phenomena} (New York, Academic).

\item""\hsk Ya. B. Zeldovich and I. D. Novikov, 1967, Soviet Astr. AJ. {\bf 10}, 602.

\endroster

\newpage
{\bf Figures}
 
FIGURE (1). This shows the form of V(z) for the exact Friedmann, static and
(non-physical) Kantowski-Sachs solutions in the $\alpha=1/3$ case. Also
shown are the curves corresponding
to $M=0$ (solid), the sonic lines $|V|=1/\sqrt{\alpha}$ (broken) and the
range of values of z (bold) in which curves can cross the sonic line
regularly.
 
FIGURE (2). This shows the form of the scale factor S(z) and the velocity
function V(z) for the
full family of spherically symmetric similarity solutions with $\alpha
=1/3$.  The exact Friedmann, Kantowski-Sachs and static solutions are
indicated by the bold lines. 
Also shown (for different values of $E_\infty$) are the
asymptotically Friedmann solutions and (for different values of $E_\infty$ and D)
the asymptotically quasi-static solutions. The latter contain a naked
singularity when the minimum of V is below 1.  
The negative V region is occupied by the asymptotically Kankowski-Sachs
solutions, though these may not be physical since the mass is negative.
Solutions which are irregular at the sonic point are shown by broken lines.
The dotted curve corresponds to a negative mass solution.

FIGURE (3). This shows the forms of the scale factor $S(z)$ and the
velocity function $V(z)$ for the asymptotically Friedmann solutions
with different values of $E_\infty$. The $z>0$ ($z<0$) solutions
contain black (white) holes for $E_*<E_\infty < E_{crit}$.

FIGURE (4). This shows the forms of the scale factor $S(z)$ and the
velocity function $V(z)$ for the asymptotically quasi-static solutions
with different values of $E_\infty$ and D. The solutions
contain a naked singularity for $E_*(D) < E_\infty(D) < E_{crit}(D)$.

\enddocument